\documentclass[preprint2,longabstract]{aastex}
\usepackage{times}
\usepackage{graphicx}

\begin{document}

\bibliographystyle{apj}

\title{The Gemini Deep Deep Survey: VIII. When Did Early-type Galaxies Form?}

\author{\medskip Roberto G. Abraham, Preethi Nair}
\affil{
   Department of Astronomy \& Astrophysics,
   University of Toronto, 50 St. George Street,
   Toronto, ON, M5S~3H4.
}

\author{Patrick J. McCarthy}
\affil{
 Observatories of the Carnegie Institution of Washington,
 813 Santa Barbara Street,
 Pasadena, CA 91101.
}

\author{Karl Glazebrook\altaffilmark{\dag}}
\affil{
  Department of Physics \& Astronomy ,
  Johns Hopkins University,
  3400 North Charles Street,
  Baltimore, MD 21218-2686.
}
\altaffiltext{\dag}{Present address: Centre for Astrophysics and Supercomputing,
Swinburne University of Technology,
1 Alfred St, Hawthorn, Victoria 3122, Australia}

\author{\medskip Erin Mentuch}
\affil{
   Department of Astronomy \& Astrophysics,
   University of Toronto, 50 St. George Street,
   Toronto, ON, M5S~3H4.
}

\author{Haojing Yan}
\affil{
 Observatories of the Carnegie Institution of Washington,
 813 Santa Barbara Street,
 Pasadena, CA 91101.
}

\author{Sandra Savaglio}
\affil{Max-Planck-Institut
  f\"ur extraterrestrische Physik, Garching, Germany}

\author{David Crampton, Richard Murowinski} 
\affil{Herzberg Institute of Astrophysics, 
National Research Council, 
5071 West Saanich Road, Victoria,
British Columbia, V9E~2E7, Canada.} 

\author{Stephanie Juneau}
\affil{
Department of Astronomy/Steward Observatory,
University of Arizona,
933 N Cherry Ave., Rm. N204,
Tucson AZ 85721-0065
}

\author{Damien Le Borgne\altaffilmark{\S}, R. G. Carlberg}
\affil{
   Department of Astronomy \& Astrophysics,
   University of Toronto, 60 St. George Street,
   Toronto, ON, M5S~3H8.
}
\altaffiltext{\S}{Present address: DSM/DAPNIA/Service d'Astrophysique, CEA/SACLAY, 91191 Gif-sur-Yvette Cedex, France}

\author{Inger J{\o}rgensen, Kathy Roth}
\affil{
  Gemini Observatory, 
  Hilo, HI 96720
}

\author{Hsiao-Wen Chen\altaffilmark{\ddag}}
\affil{
  Center for Space Research, 
  Massachusetts Institute of Technology, 
  Cambridge, MA 02139-4307
}
\altaffiltext{\ddag}{Present address: Dept. of Astronomy \& Astrophysics, University of Chicago, 5640 S. Ellis Ave, Chicago, IL 60637}

\author{Ronald O. Marzke}
\affil{
  Dept. of Physics and Astronomy,
  San Francisco State University,
  1600 Holloway Avenue, 
  San Francisco, CA 94132 
}

\begin{abstract} 
We have used the Hubble Space Telescope's Advanced Camera for Surveys \citep{for03} to
measure the cumulative mass density in morphologically-selected early-type
galaxies over the redshift range $0.8<z<1.7$. Our imaging data set covers four well-separated
sight-lines, and is roughly
intermediate (in terms of both depth and area) between the GOODS/GEMS imaging data, and the
images obtained in the Hubble Deep Field campaigns. Our images contain
144 galaxies with ultra-deep spectroscopy obtained as part of the Gemini Deep Deep Survey.
These images have been analyzed using a new purpose-written morphological analysis code which
improves the reliability of morphological classifications by adopting a `quasi-Petrosian'
image thresholding technique.
We find that at $z\sim1$ about 80\% of the stars living in the most massive galaxies
reside in early-type systems. This fraction is similar to that
seen in the local Universe. However, we detect very rapid evolution in this 
fraction over the range $0.8<z<1.7$, suggesting that over this redshift range 
the strong morphology-mass relationship seen in the nearby Universe is
beginning to fall into place. By comparing our images to published 
spectroscopic classifications, we show that little ambiguity exists in connecting spectral classes
to morphological classes for spectroscopically quiescent systems.  
However, the mass density
function of early-type galaxies
is evolving more rapidly than that of spectroscopically quiescent systems,  which we take as further
evidence that we are witnessing the formation of massive early-type galaxies over the $0.8<z<1.7$
redshift range.
\end{abstract}

 
\section{INTRODUCTION}

\label{sec:introduction}

The study of galaxy formation and evolution is one of the most active interfaces between observation and theory in contemporary astrophysics. Sophisticated numerical simulations have elucidated the key role of dark matter in driving the formation of structure \citep[e.g.][]{spr05}, while the most recent generation of simulations incorporate gas dynamics and highlight the role of feedback \citep[e.g.][]{cro06,gov06}. A number of recent galaxy surveys allow one to trace the evolution of stellar systems from high redshift to their present-day counterparts \citep[e.g.][]{wol04,bor06,fon06,sco06a}. The combination of deep imaging surveys with the Hubble Space Telescope (HST) and ground based spectroscopic surveys play a special role, as they provide both structural information and stellar content for galaxies to early epochs  \citep{bri98,sco06b}. 

Large area shallow surveys (e.g. SDSS) have provided a detailed census of galaxies in the local volume by luminosity, stellar mass, and stellar content \citep{kau03,nac03,tre04,man06}. There is a remarkably strong correlation between the structural properties of galaxies, as quantified by the Hubble type for example, stellar content as revealed by integrated spectra, and stellar mass. Galaxies with stellar masses $> 3 \times 10^{10}$M$_{\odot}$ have colors and spectra indicative of primarily passive evolution, while lower mass galaxies have blue colors and spectral features indicative of on-going star formation \citep{kau03}. The red sequence galaxies are spheroid dominated, the blue sequence galaxies are primarily disk and irregular systems. Understanding the origin of the tight connection between spectral and morphological classes and their mass dependence is critical to a complete picture of galaxy formation and evolution.

Intermediate depth surveys with HST combined with ground-base spectroscopy have shown that the Hubble sequence is largely in place at $z \sim 1$ \citep[e.g.][]{lil95,abr96,bri98,con05,kaj05}. At higher redshifts, deep surveys (e.g. GOODS) reveal galaxies with structures that are not easily classified into disks and spheriods \citep{rav06,lot06,pap05}. This trend is even stronger in small area ultra-deep surveys e.g.the Hubble Deep and Ultra-Deep Fields) \citep{abr96,yan04,coe06}, although some large disks were present at $z \sim 2.5$ \citep{lab03,sto04}.  Red and Near-IR surveys have shown that the red sequence of massive galaxies was in place at $z \sim 1$, although with number densities lower than today by a factor of $\sim 1.5 - 3$ \citep{bel03,che02,fab05}. Thus it appears that the galaxies acquired their present day morphologies in the $1 < z < 3$ epoch.

The onset of the strong correlation between morphology and stellar content and its dependance on mass is not as well constrained. One of the primary goals of the Gemini Deep Deep Survey \citep[GDDS;][]{abr04} was to use stellar-mass-selected samples to probe galaxy evolution in the critical $1 < z < 2$ range. The GDDS and other samples \citep[e.g. K20;][]{fon04} have shown that the total stellar mass density evolves mildly for $z < 1.5$ and that the high-mass end in particular is slowly evolving \citep{bri00,dic03,bel04,gla04}. Many of the most massive galaxies at $z < 2$ have red optical-to-near-IR colors \citep{gla04,pap06} and spectra dominated by old stellar populations \citep{cim04,mcc04}. In this paper we examine the evolution of the stellar mass density as a function of morphological and spectral types using the GDDS spectra and deep HST/ACS imaging. Our analysis is aided by the use of quantitative morphological measurements obtained using a new `quasi-Petrosian' threshold technique. This allows us to derive robust morphologies over a range of redshifts and apparent magnitudes. We show that the correlation between morphology and spectral class  remains strong for passively evolving objects at $z \sim 1.5$ and that the mass densities in spheroids (defined by morphology) and passive systems (defined by stellar content) evolve steeply in the $1 < z < 1.5$ range. These results strongly suggest that the core formation epoch for massive spheroids is drawing to a close in this interval. Signatures of major mergers in the passive population appear to have faded by $z \sim 1.5$ or higher, and subsequent mass evolution at $z < 1$ apparently does not significantly disturb the spectral-morphology correlation. In a subsequent paper we will extend these results to $z \sim 2$ by using NICMOS to explore the rest-frame visible morphologies of the most distant passively evolving galaxies.

We describe our GDDS+HST/ACS sample in Section 2, our approach to quantitative morphological classification in Section 3. Our results are presented in Sections 4 and 5, and discussed in
Section 6. Our conclusions are summarized in Section 7. An Appendix at the end of the paper
presents a detailed analysis of the robustness of our morphological classification technique. 
Throughout this paper we adopt a cosmology with $H_0$=70 km/s/Mpc,
$\Omega_M=0.3$, and $\Omega_\Lambda=0.7$. The Vega magnitude system is used throughout. 

\section{SAMPLE}

The GDDS is a spectroscopic survey of an optical and near-IR-selected sample targeting massive galaxies at $0.8 < z < 2.0$.  The motivation for the survey,  along with selection functions, sampling weights, details of the observations and catalogs, are presented in Paper I \citep{abr04}. Only a brief overview is given here. 

The GDDS sample is drawn from four fields in the one square-degree Las Campanas IR imaging survey \citep{mcc01,che02}, spanning a total area of 121 square arcmin. Broad-band colors were used to pre-select galaxies likely to be in the $0.8 < z < 1.7$ redshift range, and the spectroscopic component of the survey emphasized obtaining
high quality spectra for red galaxies at these redshifts. 
Very long exposures, using the `Nod \& Shuffle' technique \citep{gla01} to significantly improve
subtraction of the night sky emission lines, yielded high quality spectra from
which redshifts could be derived for 308 galaxies to a limit
 of $I_{\rm Vega}=24.5$ mag. The spectra define a one-in-two
  sparse sample of the reddest and most luminous galaxies near the
  $I-K$\, vs. $I$\, color-magnitude track mapped out by passively
  evolving galaxies in the redshift interval $0.8<z<1.7$.  This sample
  is augmented by a one-in-seven sparse sample of the remaining
  high-redshift galaxy population.  The spectra go deep enough to allow redshifts to
be obtained for $L_\star$ galaxies irrespective of star-formation
history at $z\sim 1.5$, and the survey thus targets a mass-limited sample
out to this redshift. 
Because of the importance of cosmic variance, the GDDS fields
were
carefully chosen to lie in regions of the sky where the number of red galaxies
is near the global average (in order to avoid obvious
over-densities and obvious voids).

Forty five orbits of imaging observations with the Hubble Space Telescope's
Advanced Camera for Surveys were obtained from August 2003 to June 2004.
These orbits were distributed
over 7 pointings with integration times ranging from 11.7ks to 16.4ks (per pointing).  
In areas of overlap between pointings (which occur in our SA02 and SA22 fields), the effective integration time can be much larger than this (32.6ks in SA02 and 36.1ks in SA22),
albeit over a small area. Our
rationale for obtaining relatively long integration times relative to those obtained
by surveys
such as GOODS \citep{gia04} and GEMS \citep{rix04} will be given in the next section.
A tabular summary of our observations is presented in Table~1.  ACS images of starforming
and post-starburst objects from this sample have already been presented
in earlier papers in this series \citep{sav05,leb06}, while the sample as a whole
is examined in the present paper. 

\begin{figure*}[htbp]
\begin{center}
\includegraphics[width=2.75in]{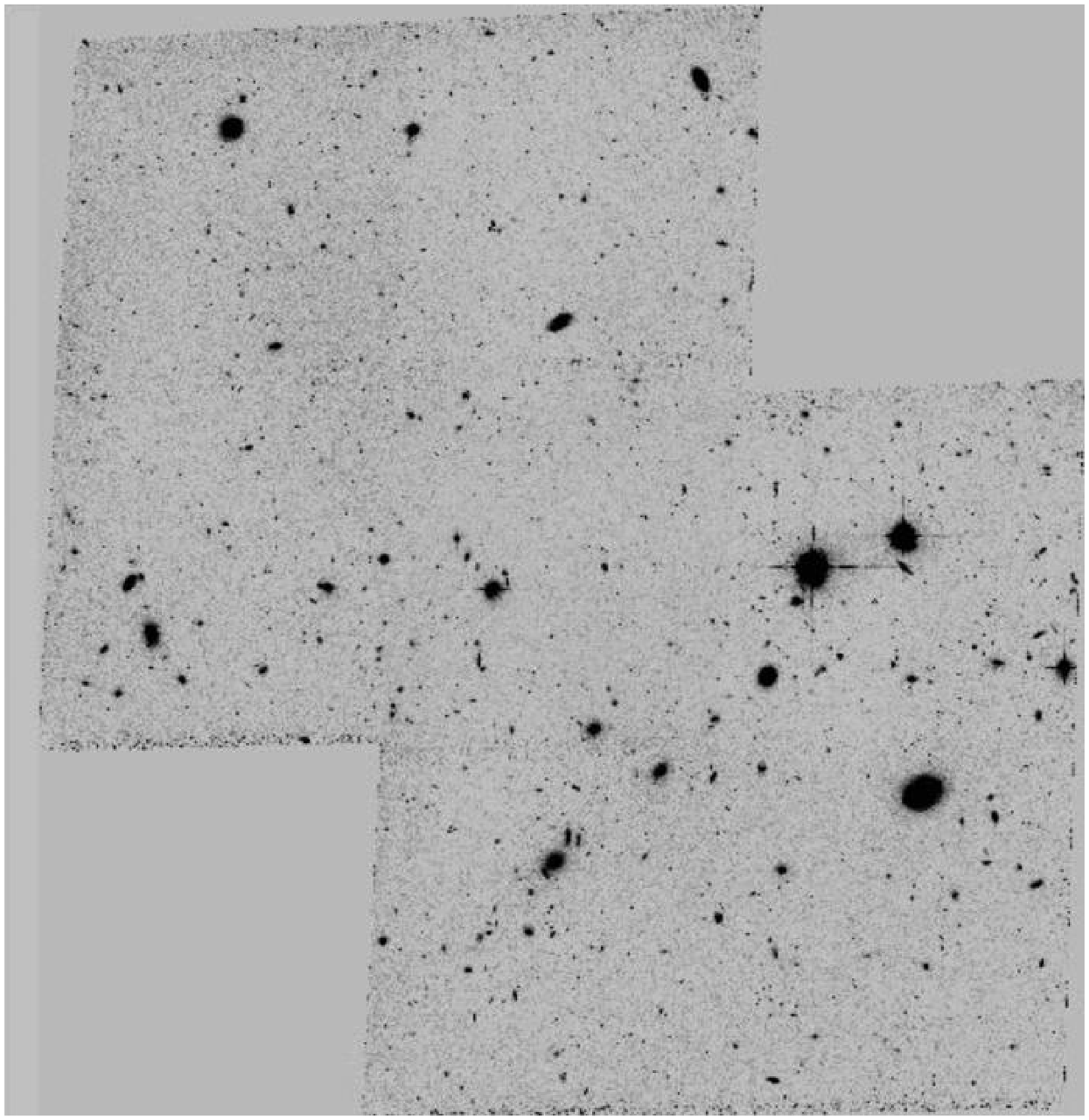} 
\includegraphics[width=2.75in]{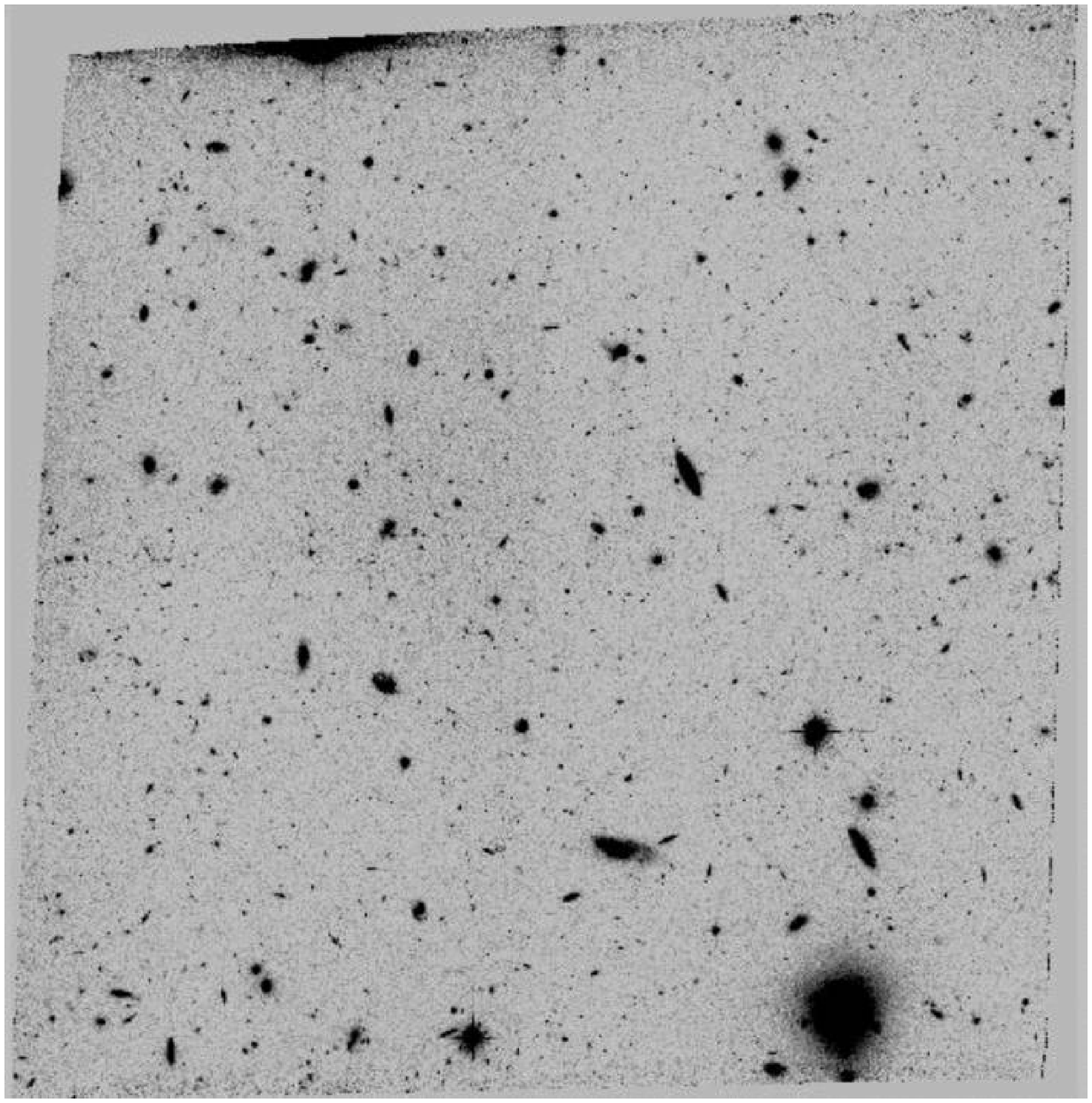} 
\includegraphics[width=2.75in]{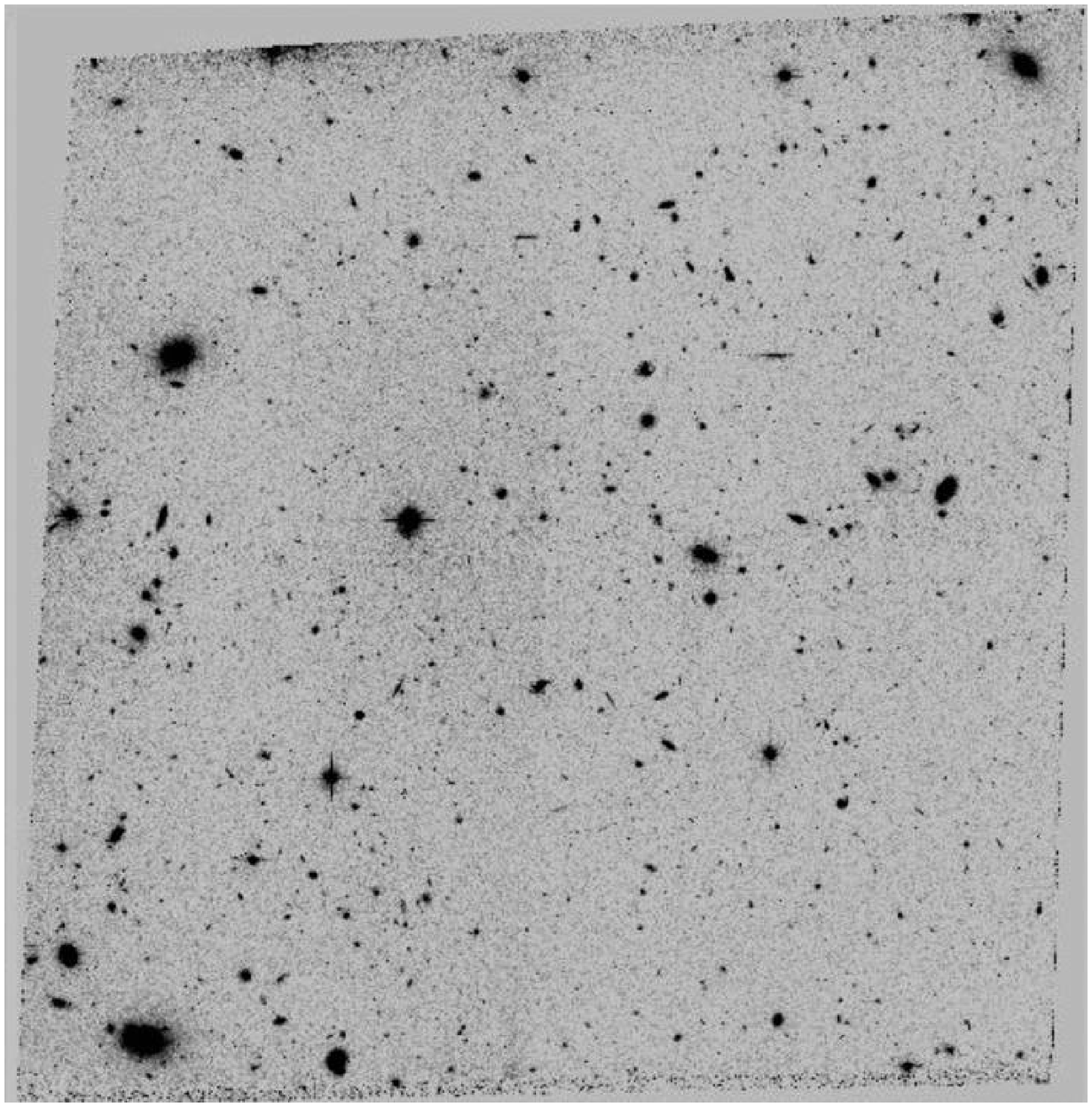} 
\includegraphics[width=2.75in]{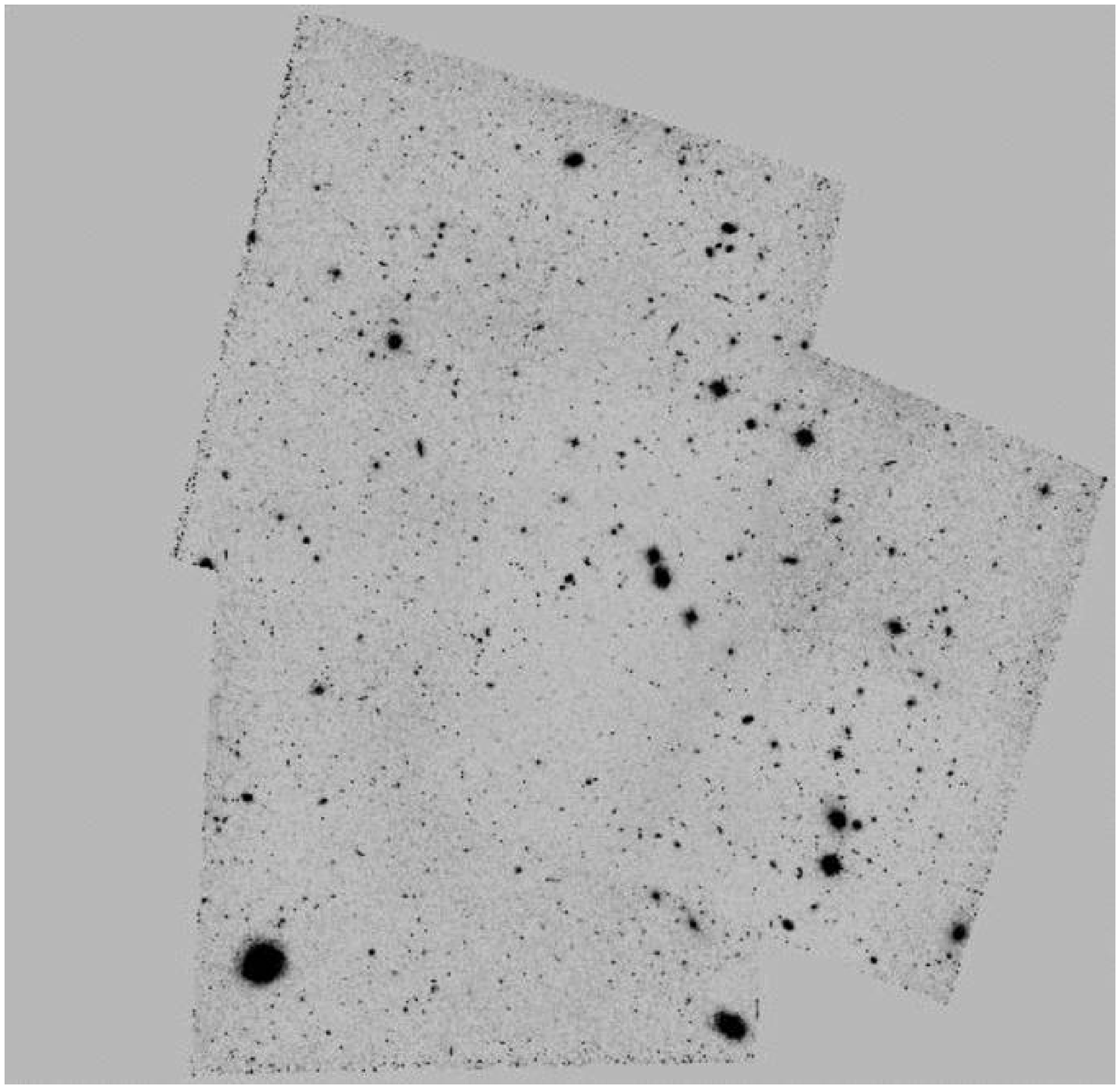} 
\caption{\label{fig:images} 
Hubble Space Telescope Advanced Camera for Surveys images showing
our SA02 field (top left),
SA12 field (top right), SA15 field (bottom left) and SA22
field (bottom right). As described in the text, the tiling
of the field pointings were chosen to maximize the integration time
on galaxies with known high redshifts. Details are given in Table~1.
}
\end{center}
\end{figure*}

Our imaging data cover 67 square arcmin, 
or 55\% of the total area of the GDDS, but by carefully choosing the locations of
the pointings within the GDDS fields, and by varying the number of pointings per field,
we were able to obtain ACS images for
63\% of the galaxies in our spectroscopic sample. 
Fully reduced ACS F814W images for each of the four fields are shown in Figure~\ref{fig:images}. Our
field orientations were defined to produce additional depth for high priority targets within the allocated orbit total. This resulted in an unusual and non-uniform coverage within each field. 
The objects within our GDDS+ACS sample are representative of the parent distribution in their local densities and distribution on the sky. Our sub-sample of GDDS galaxies imaged with ACS are neither drawn from unusually dense regions nor from low density voids. 

\begin{deluxetable}{cccccc}
\tablecaption{Log of observations\label{tab:obs}}
\tablecolumns{6}
\tablewidth{0pc}
\tabletypesize{\small}
\tablehead{
  \colhead{Pointing} &
  \colhead{RA} &
  \colhead{Dec} &
  \colhead{\# Visits} &
  \colhead{\# Orbits} &
  \colhead{Integration time}\\
  \colhead{} &
  \colhead{(J2000)} &
  \colhead{(J2000)} &
  \colhead{} &
  \colhead{} &
  \colhead{(s)}
}
\startdata
\cutinhead{\em SA02 (2 pointings, 14 orbits, 32676s total integration time, 19.9 arcmin$^2$ area)}
1 & 02:09:43.48 & -04:36:42.0 & 3 & 7 & 16338 \\
2 & 02:09:37.43 & -04:38:34.6 & 3 & 7 & 16338 \vspace{0.5cm} \\
\cutinhead{\em SA12 (1 pointing, 6 orbits, 14640s total integration time, 11.5  arcmin$^2$ area)}
1 & 12:05:26.71 & -07:23:34.2 & 2 & 6 & 14640 \vspace{0.5cm} \\
\cutinhead{\em SA15 (1 pointing, 5 orbits, 12200s total integration time, 11.4  arcmin$^2$ area)}
1 & 15:23:50.38 & -00:04:34.3 & 2 & 5 & 12200 \vspace{0.5cm} \\
\cutinhead{\em SA22 (3 pointings, 5 orbits, 36070s total integration time, 24.0  arcmin$^2$ area)}
1 & 22:17:38.15 & +00:16:35.1 & 2 & 5 & 11670 \\
2 & 22:17:37.71 & +00:14:21.6 & 2 & 5 & 12200 \\
3 & 22:17:45.09 & +00:15:33.2 & 2 & 5 & 12200
\enddata
\end{deluxetable}

\begin{deluxetable}{cccccc}
\tablecaption{Summary of the ACS imaging sample\label{tab:summary}}
\tablecolumns{6}
\tablewidth{0pc}
\tabletypesize{\small}
\tablehead{
  \colhead{Sample} &
  \colhead{SA02} &
  \colhead{SA12} &
  \colhead{SA15} &
  \colhead{SA22} &
  \colhead{All}
}
\startdata
Galaxies in the LCIRS parent sample & 998 & 839 & 748 & 1011& 3596\\
LCIRS galaxies with $18<I<24.0$ mag & 375 & 287 & 315 & 542 & 1519\\
Galaxies in the GDDS sample & 48 & 29 & 26 & 90 & 193\\
GDDS galaxies with $18<I<24.0$ mag & 39 & 24 & 11 & 70 & 144\\
\enddata
\end{deluxetable}

%

Using the methodology described in the next section,
reliable morphologies can be obtained for galaxies
down to $I_{F814W}=24.0$ mag. 
Our images contain 3596 galaxies in the LCIRS parent sample, 1519 of
which are brighter than our morphological analysis limit. Of these, 144 galaxies are in the GDDS.
A detailed breakdown of the numbers of objects in individual fields is
given in Table~2. 

\section{METHODOLOGY}

The approach to quantitative morphological classification adopted in the
present paper is based on an updated version of the now fairly well-established technique of
subdividing the galaxy population into classes on the basis of position 
on an asymmetry vs. concentration diagram.
This approach was first used in  \citet{abr96} to analyze the
Hubble Deep Field, and variations of it
have now been widely adopted \citep{bri98,con03,lot04,sca06,lis06}. While 
details of implementation differ amongst authors, the general approach has been shown to work well
for subdivision into broad classifications (e.g. early-types vs. spirals vs. 
peculiars/mergers). Unfortunately the more subtle distinctions between the 
morphological classes (e.g. elliptical galaxies vs. S0 galaxies) are not
captured by the system. 

Our analysis in this paper is based on 
a new publicly-available code,
$\tt MORPHEUS$, which produces a large number of measurements
which can be used to constrain the morphological properties of galaxies
in addition to concentration and asymmetry\footnote{{\tt MORPHEUS} can
be downloaded from the following web site:
http://odysseus.astro.utoronto.ca/$\sim$abraham/Morpheus. Please note that
the code base is
evolving and the version
used in the present paper is the July 2006 release. In addition to incorporating some relatively new parameters (e.g. the Gini coefficient
from Abraham et al. 2003 and M20 from Lotz et al. 2004) {\tt MORPHEUS} also
incorporates improvements suggested by others for ways to better measure well-established
parameters such as asymmetry (e.g. Conselice et al. 2003).}. The optimal
set of parameters for characterizing galaxy morphology will be explored in 
Nair et al. 2007 (in preparation), but as will be shown below, 
for the simple purpose of isolating early-type galaxies
from all other galaxy types, a straight-forward diagram based on only two parameters,
namely the Gini coefficient and asymmetry, works rather well.  
The Gini coefficient is fairly new to astronomy  (being introduced
in \citet{abr03} and improved in \citet{lot04}), but it has been
used in econometrics for nearly a century to quantify the inequality of wealth distributions in human populations \citep{gin12}. When applied to galaxy images, the Gini coefficient provides a quantitative measure of the inequality with which a galaxy's light is distributed amongst its constituent pixels, and it can be used as a sort of 
generalized concentration index that does not depend on galaxy symmetry. Furthermore,
as shown the Appendix~A, the Gini coefficient remains a surprisingly robust statistic
even in the face of 
morphological $K$-corrections.
We therefore consider classification based on position in the Asymmetry-Gini diagram (referred to
as the A-G diagram, or A-G plane) 
to be 
an updated and improved version of classification on the basis of the Asymmetry-Concentration
diagram \citep{abr96}. The A-G plane will be the central diagnostic diagram in the present paper
for purposes of morphological classification.

As noted earlier,  \citet{lot04} presented a valuable refinement of the original
definition of the Gini coefficient given in \citet{abr03}. \citet{lot04} computed
the Gini coefficient using circular apertures scaled to multiples of the Petrosian radius. 
In Section~\ref{sec:quasi} below we will show how this idea can be extended
using so-called `quasi-Petrosian'  isophotes, which retain all improvements to the Gini coefficient 
presented by \citet{lot04}, and improve the statistic further so it works better for galaxies 
of arbitrary shape. But before describing our improved statistic, we will describe
the importance of having data deep enough to be worth analyzing with these
improved tools. Further discussion can be found in Appendix~A.

\subsection{Required depth}

The long integration times obtained in our ACS fields were
motivated by our need for reliable morphological classification of
early-type galaxies  out to $z\sim2$. Morphological
$K-$corrections impact early-type galaxies in a qualitatively different 
manner
from how they impact late-type galaxies (see \citet{abr96} and \citet{bri98} for a more detailed discussion). Late-type galaxies are dominated
by irregular knots of star-formation which become more prominent as the 
observed wavelength probes further to the blue, typically leading
to increased asymmetry. On the other hand, early-type galaxies are (by
definition) smooth and rather symmetric, and the changes that occur 
are more uniform --- the central parts of the galaxy remain brightest at
all redshifts, and the asymmetry remains low.  As is shown in greater
detail in Appendix~A, morphological
classification of early-types from a single band of observation
is possible out to
high-redshifts, {\rm provided} the 
data go deep enough to probe the galaxies out to similar physical radii
over the redshift range of interest. Obtaining data of sufficient depth
is the crucial ingredient --- probably the greatest source of confusion in morphologically classifying early-type galaxies at high redshifts is systematic misclassification of early-intermediate spirals as E/S0 systems, because at high-redshifts $(1+z)^4$ flux diminution lowers the surface brightness of disks below the threshold of visibility, thus increasing the prominence of the central bulge. Our ACS integration times were chosen to
be deep enough to avoid this effect.

\begin{figure*}[htbp]
\centering
\includegraphics[width=6.5in]{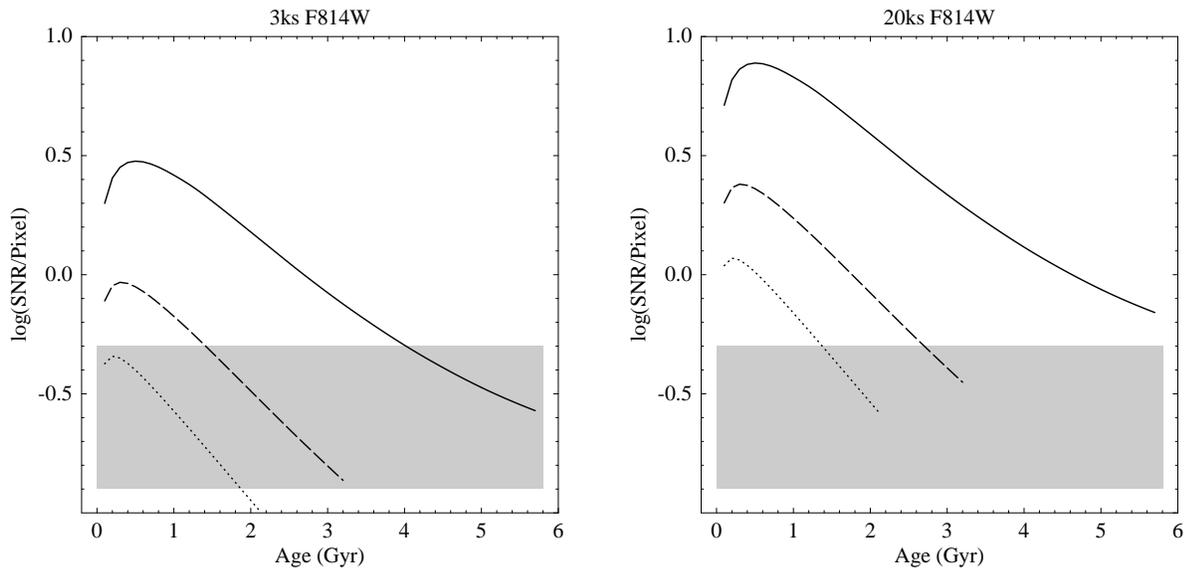} 
\caption{ 
Predicted F814W-band $\log_{10}$(signal-to-noise/pixel) as a function
of age for a
reference stellar population seen with the HST ACS camera at 
$z=0.7$ (solid lines), $z=1.2$ (dashed lines),
and $z=1.7$ (dotted lines).  The panel on the left corresponds to an exposure time of 3ks, 
and the panel on the right corresponds to a 20ks integration. 
Calculations are based on a stellar population with a surface stellar
mass density of $10^8 M_\odot$/kpc$^2$, which is comparable to that of the
solar neighborhood. When applying K-corrections we assumed
an exponential star-formation history with an e-folding timescale of 1 Gyr, 
a Miller-Scalo IMF, and 30\% solar metallicity. To approximately
account for the light smoothing
and connected-component optimizations 
done by SExtractor when segmenting galaxy pixels from the background sky, 
we assumed an effective pixel scale of 0.1 arcsec/pixel, and that a SNR/pixel 
greater than 0.5 is detectable. (Note that $\log_{10}(0.5)=-0.301$). With these assumptions,
the grey regions shown in the plot are unobservable due to low 
signal-to-noise. Another factor impacting the observability of the stellar
populations is the age 
of the Universe, so the curves shown are limited to ages less than that of the 
Universe at each epoch of observation. See text for details. }
\label{fig:SNR}
\end{figure*}

Figure~\ref{fig:SNR} shows the signal-to-noise/pixel as a function of age and
redshift for HST ACS observations of a projected stellar mass density corresponding to that of the
Milky Way at the solar radius. The panel on the left corresponds to an exposure time of 3ks, 
and the panel on the right corresponds to a 20ks integration. Different lines correspond to
observations of galaxies at $z=0.7$, $1.2$, and $1.7$ (the range of redshifts explored
in the present paper).
The logarithm of signal-to-noise/pixel is shown as a function of the age of the stellar
population, assuming the fairly typical stellar population model described in the caption. To
account (rather crudely) for the light smoothing
and connected-component optimizations 
done by the pre-processing imaging segmentation
software used (discussed further below),
we assume an effective pixel scale of 0.1 arcsec/pixel, and that a SNR/pixel 
greater than 0.5 is detectable\footnote{This number is somewhat arbitrary, because in reality the limiting isophotal signal-to-noise is dependant on the  smoothing parameter and connected-component algorithm (i.e. 4-point vs. 8-point connectivity) adopted by the segmentation software used to separate out a galaxy's pixels from the underlying sky. A value around 0.5 is reasonable assuming fairly heavy smoothing and 4-point connectivity.}. On this basis, Figure~\ref{fig:SNR} shows that at $z\sim0.7$ one is
able to probe out to large radii in the rest-frame of galaxies even with
relatively shallow (e.g. 1 orbit, or 3ks) optical observations obtained with HST. At this redshift, projected mass
densities corresponding to the intermediate regions of large disk galaxies are
observable for around 60\% of the maximum possible age for the galaxy (defined
as the age of the Universe at the epoch of observation). However, beyond this
redshift biases quickly become rather severe --- structural information at optical wavelengths comes only from the densest
regions in galaxies (such as bulges), or from the youngest parts of galaxies
(such as bright star-formation complexes with ages of less 1 Gyr). By $z=1.7$ it
is simply no longer possible to use shallow HST optical data to compare fairly
fundamental characteristics, such as the size and morphology of galaxies, with
those of local objects. With 20ks integration times, much fairer comparisons can be
made --- surface mass densities near that of the solar neighborhood with maximal
ages can be probed in an unbiased way out to about $z=1.2$, and at $z=1.7$
one is able to probe out to radii comparable to those probed by shallow
HST imagery at $z=0.7$. If, as suggested by Bouwens et al. 2005 (see
also Iye et al. 2007), significant star-formation
activity only begins around $z\sim6$ (when the Universe is already 
about 1 Gyr old, so that
the maximal age of a stellar population at $z=1.7$ is 2.8 Gyr instead of the 3.8 Gyr
age of the Universe at that redshift), then
Figure~\ref{fig:SNR} suggests that one would be able
to probe maximally old stellar populations at solar-neighborhood mass
densities out the limits of our survey with 20ks integrations. Our goal
in defining the rather complicated-looking ACS imaging field geometry shown in 
Figure~\ref{fig:images} was to cover as many GDDS galaxies with known
redshifts as possible, subject to the constraint of having at least 12ks of
integration everywhere, and up to $33$ks of integration in areas
where fields overlap. These areas of overlap were chosen to correspond
to regions with many high-redshift galaxies.

Having obtained data of the required depth, the next ingredient
needed for reliable morphological classification of early-type galaxies
is construction of a morphological catalog whose measurements have been synchronized (as
closely as possible) to a common physical
radius in the rest-frame of each galaxy. Standard tools can take us part of
the way to this goal, and the next stage in our analysis was to create a photometric catalog
using {\tt SExtractor} \citep{ber96}, which
constructs a photometric catalog by isolating (or
`segmenting') galaxies from the background sky on the basis
of fairly complicated criteria, based on differential thresholding followed
by stages of  image splitting based on connected-component analysis. Synchronizing
measurements to a common physical rest-frame is done by using the
{\tt SExtractor} catalog and segmentation maps as the input for our own
software ({\tt MORPHEUS}), which measures galaxy structural parameters
contained within a `quasi-Petrosian' isophote, as defined in the next section. 

\subsection{Definition of the Quasi-Petrosian Isophote}
\label{sec:quasi}

The classic prescription for dealing with limiting isophote mismatches introduced
by $(1+z)^4$ surface
brightness diminution of galaxies is to measure galaxy properties within a circular aperture
whose size is a multiple of a galaxy's Petrosian radius
\citep{pet76}. Our more
modern formulation of this basic idea  is valid for galaxies
of arbitrary shape, and is thus more generally useful
for analysis of the very diverse population of galaxies seen in
deep HST images.

The `quasi-Petrosian' isophote is constructed using an
algorithm which works for galaxies of arbitrary shape. All
pixels in the galaxy  image (defined using the {\tt SExtractor} segmentation
map) are
sorted in {\em decreasing} order of flux to construct an array, 
$f_i$, containing the flux in the $i^{\rm th}$ sorted pixel.
This array is then summed over to construct a monotonically
{\em increasing} curve of cumulative flux
values:
\begin{equation} 
\mathcal{F}_i=\displaystyle\sum_{j=1}^i f_j.
\end{equation}
In analogy with the definition of the Petrosian radius for
a circular aperture, we calculate the Petrosian isophote
by determining the pixel index $i$ 
which satisfies the following equation:
\begin{equation}
f_i = \eta\times \left( \frac{\mathcal{F}_i}{i}\right).
\label{quasiPetroEqn}
\end{equation}

\noindent 
Note that $\frac{\mathcal{F}_i}{i}$  is the cumulative mean surface brightness in the sorted array.
The flux value of the pixel which solves
the equation defines the {\em quasi-Petrosian
isophote}. All pixels brighter than $f_i$ are retained, and all pixels
fainter than $f_i$ are discarded. The parameter $\eta$ is
fixed for all galaxies in the sample, and experimentation
has shown us that $\eta=0.2$ is a good choice, probing far enough 
into the outskirts of nearby galaxies to get outside of regions dominated by
nuclear bulges, but not so far out that low and high-redshifts samples
cannot
be fairly compared. Because the data are discrete,
Equation~\ref{quasiPetroEqn} will rarely be solved exactly, but
a perfectly adequate approximation is obtained by simply
noting the index of the first zero crossing of 
$f_i - \eta\times \left( \frac{\mathcal{F}_i}{i}\right)$.  A zero 
crossing is guaranteed to occur at the last pixel index, and when
this occurs it indicates that no quasi-Petrosian isophote
exists, and the algorithm converges on the standard isophote as the best approximation
to the quasi-Petrosian isophote. We consider this a rather
graceful failure mode, and when it occurs, {\tt MORPHEUS} simply flags this condition
and proceeds with its analysis using the standard isophote.

It is interesting to compare Equation~\ref{quasiPetroEqn} with the
definition of the Petrosian radius adopted by the SDSS collaboration \citep{bla01}.
In this formulation, one starts with $I(r)$, the azimuthally averaged surface brightness profile
of the galaxy. For a given profile, the Petrosian radius, $r_p$, is the radius
$r$ at which which the following relation holds:
\begin{equation}
\frac{\int_{0.8r}^{1.25r} dr^{\prime} 2\pi r^{\prime} I(r^\prime)/[\pi (1.25^2-0.8^2)r^2]}{\int_0^r dr^\prime 2\pi r^\prime I(r^\prime)/(\pi r^2) }=0.2.
\label{realPetroEqn}
\end{equation}
Thus the Petrosian radius 
is simply the radius at which 
the local surface brightness in a circular annulus is equal to 20\% of 
the mean surface brightness within
the annulus. Understanding this, we see that while Equation~\ref{quasiPetroEqn}
looks very different from Equation~\ref{realPetroEqn}, both
operate in a similar way.  The factor $ \left( \frac{\mathcal{F}_i}{i}\right)$ 
is simply the mean surface brightness of a galaxy's brightest $i$ pixels, so Equation
\ref{quasiPetroEqn} is essentially comparing the surface brightness at a point to
a scaled surface brightness interior to an isophote --- the key idea in the
standard definition of a Petrosian radius. If a galaxy's light distribution is circularly symmetric
and monotonically decreasing from
a central pixel, then the positions of the brightest $i$ pixels will closely
describe the filling-in of a bounding circle, so in this special
case the analogy  between Equation~\ref{quasiPetroEqn}
and the standard Petrosian formalism is 
nearly exact.
However, Equation~\ref{quasiPetroEqn} remains valid for galaxies of
arbitrary shape, and is both operationally
easier to calculate and more robust than Equation~\ref{realPetroEqn}.

\begin{figure*}[htbp]
\begin{center}
\includegraphics[width=4.5in]{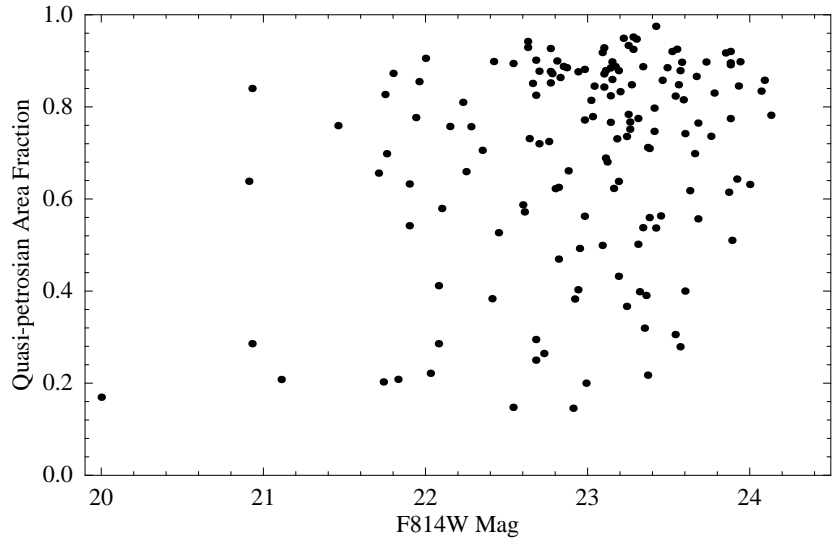} 
\caption{\label{fig:qpf}  
The quasi-Petrosian area fraction (the fraction of each galaxy's pixels above the
isophotal threshold) plotted as a function $I$-band magnitude. 
}
\end{center}
\end{figure*}

\begin{figure*}[htbp]
\begin{center}
\includegraphics[width=4.5in]{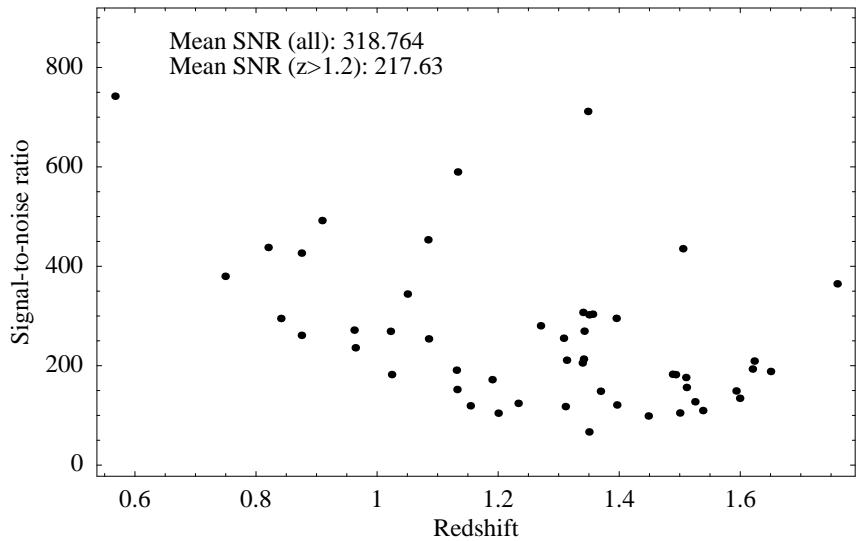} 
\caption{\label{fig:snr}  
Signal-to-noise ratio as a function of redshift for galaxies with stellar masses
$\log_{10}(M/M_\odot)>10.5$. These objects constitute our
massive galaxy subsample (see text for details). The mean signal-to-noise
level of the galaxies shown, and of the subset of objects with $z>1.2$, are noted
near the top of the figure.
}
\end{center}
\end{figure*}

Figure~\ref{fig:qpf} shows the
the fraction of each galaxy's pixels above the
isophotal threshold defined by solving Equation~\ref{quasiPetroEqn} for $\eta=0.2$ in our sample. As noted above, data
that is too shallow for reliable classification and analysis is flagged by non-convergence
of Equation~\ref{quasiPetroEqn}.
As shown
in Figure~\ref{fig:qpf}, the fraction of each galaxy's pixels above the isophotal threshold is less than unity for all galaxies in our sample --- therefore
a quasi-Petrosian isophote exists for {\em all} galaxies in our sample. 
This is not the case when our analysis includes galaxies
fainter than $I_{F814W}\sim24$ mag, and
the existence of
a quasi-Petrosian isophote can be used as a sensible metric
for determining how deep morphological measurements should be
pushed to when analyzing a given dataset\footnote{In other words,
if the quasi-Petrosian threshold is not being met for an object, then
the data is so shallow that only the inner part of the galaxy is being
probed.  We thank Edward Taylor
at Leiden Observatory for pointing out to us that meeting the
quasi-Petrosian threshold condition is a convenient way to 
determine if data is deep enough for reliable morphological
work.}.  It is interesting to
consider how a sample defined by quasi-Petrosian area fraction resembles
one based on a signal-to-noise threshold\footnote{There are several 
different definitions for the signal-to-noise of an image. In the present paper a
galaxy's
signal-to-noise means the ratio of its background-subtracted flux to the product of
the square root of its isophotal area times the standard deviation of the sky noise.}, so the signal-to-noise vs. redshift
distribution of the massive galaxy sample considered in the present
paper is shown in Figure~\ref{fig:snr}. (See below for the precise
definition of this sample). As noted on the figure,
our massive galaxy sample has a mean signal-to-noise
ratio of about 320. The subset of objects with $z>1.2$ has a mean signal-to-noise
ratio of about 220.

Figure~\ref{fig:qpf}
also illustrates a drawback of the quasi-Petrosian approach. Because only a fraction of
each galaxy's pixels are being
used in our morphological computations, some information is being lost, 
albeit the pixels with the lowest signal-to-noise in each
galaxy. The median fraction of pixels lost
is $\sim$50\% for galaxies at \hbox{F814W$\ <22\ $mag}, dropping to
$<$20\% for galaxies at \hbox{F814W$\ \sim24\ $mag}. Throwing data
away is never a happy choice to have to make, but because
of $(1+z)^4$ cosmological surface-brightness dimming, it seems
essential to do this (in a careful way) if fair comparisons of
morphology over a range of redshifts are to be undertaken. Typically,
only the high surface-brightness portions of nearer galaxies in a sample would remain
visible at the highest redshifts probed by the same sample, 
so excising the low-surface brightness portions
of bright objects is needed to harmonize regions
of comparison over a broad range of redshifts. 

We conclude this section by noting, in passing, that there is a rather pleasing conceptual connection
between our definition of the quasi-Petrosian isophote given above, and the
definition of the Gini coefficient given in \citet{abr03}. Both statistics
are obtained from simple computations on a ranked pixel list. 

\section{THE MORPHOLOGICAL MIX OF GALAXIES IN THE GDDS FIELDS}

\begin{figure*}[htbp]
\begin{center}
\includegraphics[width=6.5in]{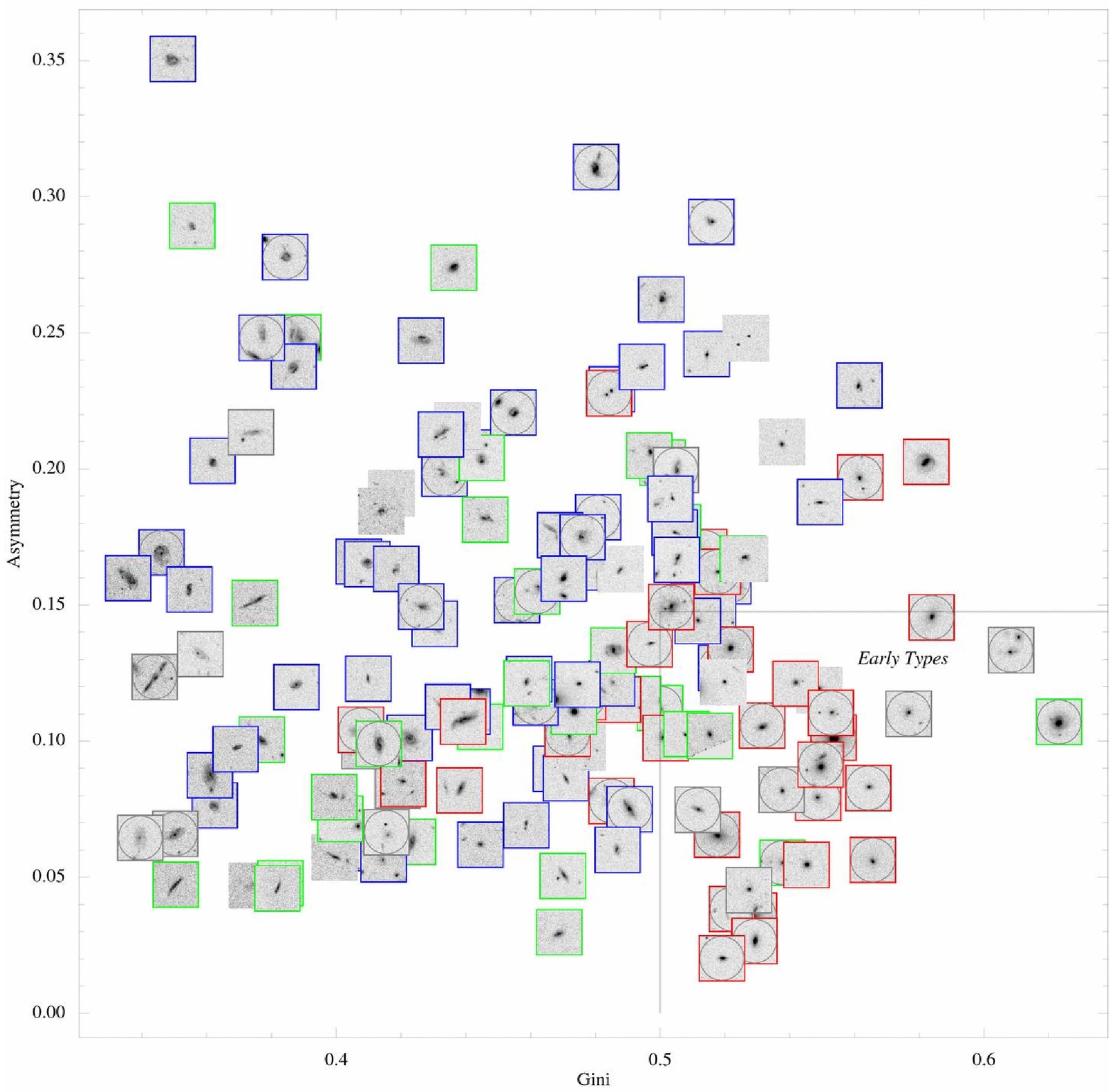} 
\caption{\label{fig:ga} 
Asymmetry vs. Gini coefficient for galaxies in the Gemini Deep Deep Survey. The
region of this diagram used to isolate early-type systems is marked at the bottom-right
corner of the figure.
Individual galaxies are shown as  5 arcsec X 5 arcsec `postage stamps'. Circled
objects are included in the $\log_{10}(M/M_\odot)>10.5$ massive galaxy sub-sample defined in the text. 
The border of each postage stamp is
is colored according to its spectral classification based on the system described
in Paper I. Objects with spectral classifications
corresponding to actively star-forming galaxies (classes 100, 110) are shown in
blue; quiescent and nearly-quiescent systems (classes 001 and 011) are shown
in red; intermediate activity systems (classes 010) are shown in green. Unclassifiable
systems with known redshifts are shown with a gray border. (These unclassifiable 
objects had very limited wavelength
coverage because they overlapped with nearby objects on our multi-object spectroscopy
masks). Systems with no
spectroscopic redshifts are shown without any border.
See text for details.
}
\end{center}
\end{figure*}

Figure~\ref{fig:ga} shows small `postage stamp' images of each GDDS galaxy, color coded by their spectral type, in the A-G plane. Galaxies with purely passive spectral types, identified with red borders, have primarily compact early-type morphologies and lie preferentially in the high Gini-low asymmetry region of Figure~\ref{fig:ga}. Galaxies with spectral types characteristic of active star formation have primarily disk and late-type morphologies and occupy a large range of the A-G plane, strongly favoring the high asymmetry and and low Gini regions. Galaxies with intermediate or composite spectral types occupy the region of the A-G plane between the passive and star forming galaxies. Figure~\ref{fig:ga} clearly shows that a strong correlation between morphology, quantified by the A-G plane, and spectral class is in place at the redshifts spanned by the GDDS. For the early-type galaxies in particular, selection by spectral class or morphology are nearly equivalent, and selection by morphology can be undertaken using simple linear cuts in the A-G plane. Throughout the remainder of this paper, we will define early-type galaxies to be those systems with ${\rm G}>0.5$ and ${\rm A}<0.15$. The choice of ${\rm G}>0.5$ is important, and its rationale is discussed in greater detail in Appendix~A. The choice of ${\rm A}<0.15$ is based simply on visual inspection of Figure~\ref{fig:ga}, but is not particularly fundamental for the limited purposes to which we are presently applying
this diagram (namely, for determining the cumulative mass function of massive early-type galaxies). This can
be seen directly from Figure~\ref{fig:ga}, where we have drawn circles
around those objects which meet the mass completeness cut used when 
determining our mass functions
(to be described in the next section). In fact,
so few high mass systems lie above the asymmetry cut that our main conclusions do not change if no asymmetry cut is imposed at all. Therefore, our classification system for massive early-type galaxies is fundamentally based on the Gini coefficient.

\begin{figure*}[htbp]
\begin{center}
\includegraphics[width=6.0in]{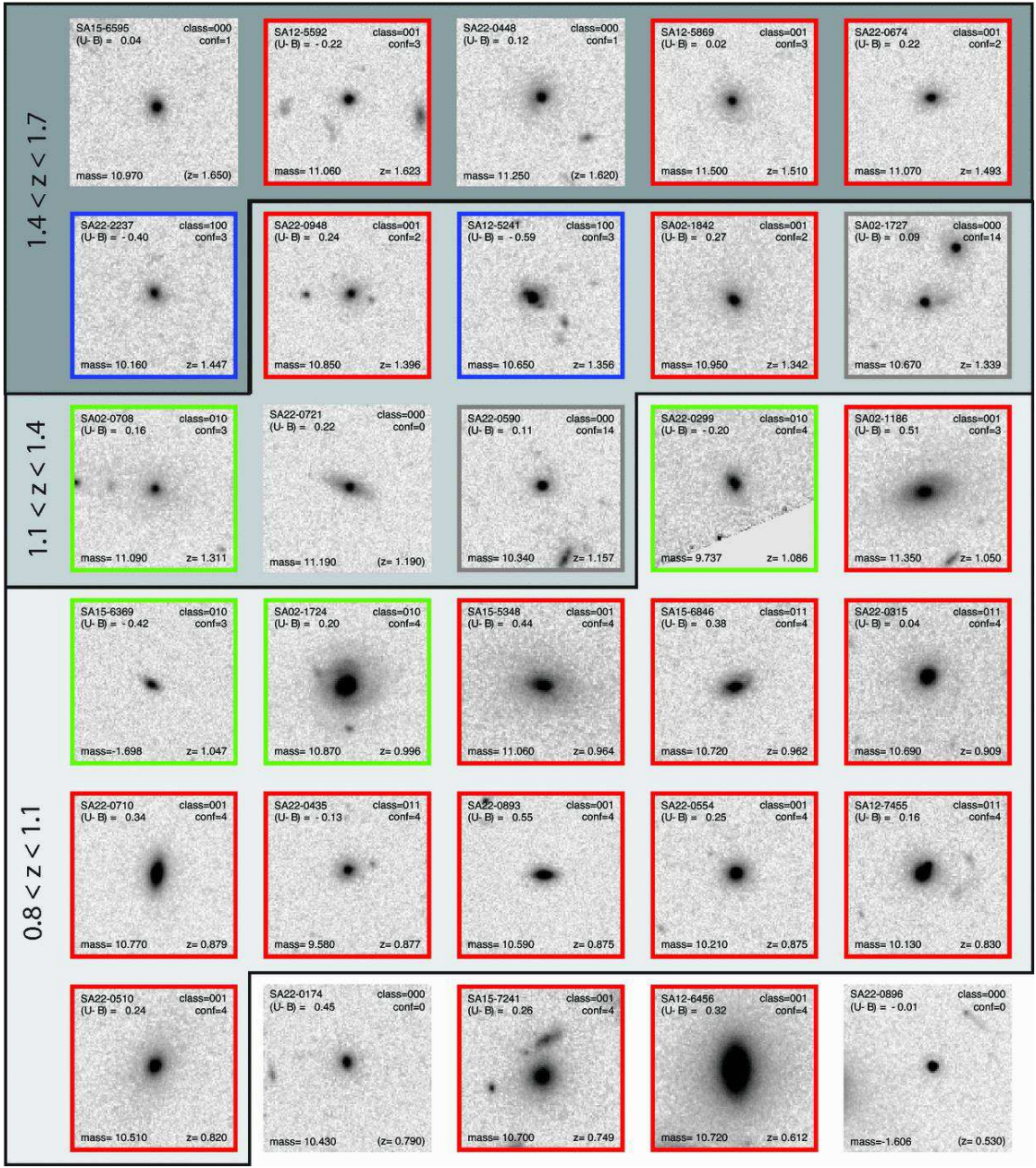} 
\caption{  
\label{fig:earlyTypeMontage}
Postage-stamp images showing the morphologies of the 30 objects classified
as early-type galaxies on the basis of the system described in the text. Galaxies are
shown sorted in order of decreasing redshift. Each image
is 5 arcsec by 5 arcsec in size, and labeled with the galaxy's ID number, spectroscopic
classification, redshift confidence class, rest-frame (U-B) color, redshift, and stellar
mass inferred from our best-fit model. Objects without high-confidence spectroscopic
redshifts have their redshifts labeled in parentheses. The borders of each image 
are colored according to spectroscopic classification, as described in the caption
to Figure~\ref{fig:ga}. Gray regions surrounding
groups of postage stamps indicate which of three broad
redshift bins 
the objects fall within. These bins are used to calculate the
cumulative stellar mass function described in \S\ref{sec:mfunc}.}
\end{center}
\end{figure*}

The concordance between morphology and spectral type is shown more clearly in Figure~\ref{fig:earlyTypeMontage},
where we present the individual images of
galaxies classed as being early-type using our formalism. Visual 
inspection of this figure is consistent with all galaxies being early-type systems.
Approximately 15\% (5/30) of the galaxies do show weak disks, but without spiral structure,
and these are probably S0 or S0/a systems. 
Clearly the A-G diagram does not discriminate between E and S0 galaxies.

\section{THE EVOLVING MASS-DENSITY FUNCTION OF MASSIVE EARLY-TYPE GALAXIES}
\label{sec:mfunc}

In this section we will attempt to synthesize the evolutionary history of
massive early-type galaxies by exploring the growth in their cumulative mass
density in three broad redshift bins. The mass density functions
were computed using the standard
$V_{max}$ formalism described in
Paper III \citep{gla04}, to whom the reader is referred
for details.
For present purposes it suffices to just highlight a few points that
are important to bear in mind
when interpreting the models.

It is important to note that our models assume a
 \citet{bal03} Initial Mass Function (IMF), which has the same
high-mass slope as a Salpeter IMF
but which has a break at 1 solar mass (providing more realistic $M/L$
values)\footnote{The following convenient relationship (accurate to within a
few percent independently of the SFH) can be used to convert between
mass-to-light ratios measured using our chosen IMF and those
determined with a Salpeter IMF: $M/L_K (BG) = 0.55 M/L_K(SP)$.}.
Whenever we compare our results
to those in the literature, we always convert
to this standard IMF. 
Typical uncertainties in our
model masses are around 0.2 dex in the main $K<20.6$ sample considered
in the present paper.

Because the GDDS is a sparse-sampled survey, each galaxy in the
survey acts as a proxy for a number of other objects with similar magnitudes and
colors, so determining the stellar mass contributed
by an individual galaxy  is only the first step in
computing the mass function. The volume-weighted contribution of this
galaxy to the integrated
mass density in a redshift bin is computed using the sampling weights
tabulated in Paper~I. These weights account for the fact that the GDDS is not 
completely homogeneous, because when designing spectroscopic masks a higher priority
was given to red galaxies than to blue galaxies. The sampling weights 
quantify the selection
probability as a function of $I-K$ and $K$, with reference to
the full wide-area LCIRS tiles which formed the basis for the
GDDS sample (and so also account approximately for
the effect of cosmic variance). 

For maximum robustness in the present paper
we focus our analysis on cumulative mass functions for systems above a mass threshold. 
We adopt a threshold of
$\log_{10}(M/M_\odot)>10.5$. As shown in Figure~1 of Paper III,
this is the limit to which the GDDS is mass-complete for $z<1.7$. In other words, red
galaxies with stellar masses of $\log_{10}(M/M_\odot)<10.5$
would begin to drop out of our survey at the highest redshifts. 	Another
issue that arises when computing the cumulative mass function is 
the treatment of objects without redshift information. Because the GDDS
has a redshift completeness of over 85 percent in the redshift
range explored in this paper, the issue is not
of overwhelming importance, and none of the conclusions in this paper
change if galaxies without redshifts are simply omitted. However,
for the sake of consistency we will treat these objects in the same way as they were
treated in Paper~III. We therefore assign photometric redshifts (computed
for the LCIRS by \citet{che02}) to objects without
redshifts, and to objects whose redshift 
confidence class is $<2$. (See Table~3 of Paper I for a detailed description of these confidence
classes).  

\begin{figure*}[htbp]
\begin{center}
\includegraphics[width=5.6in]{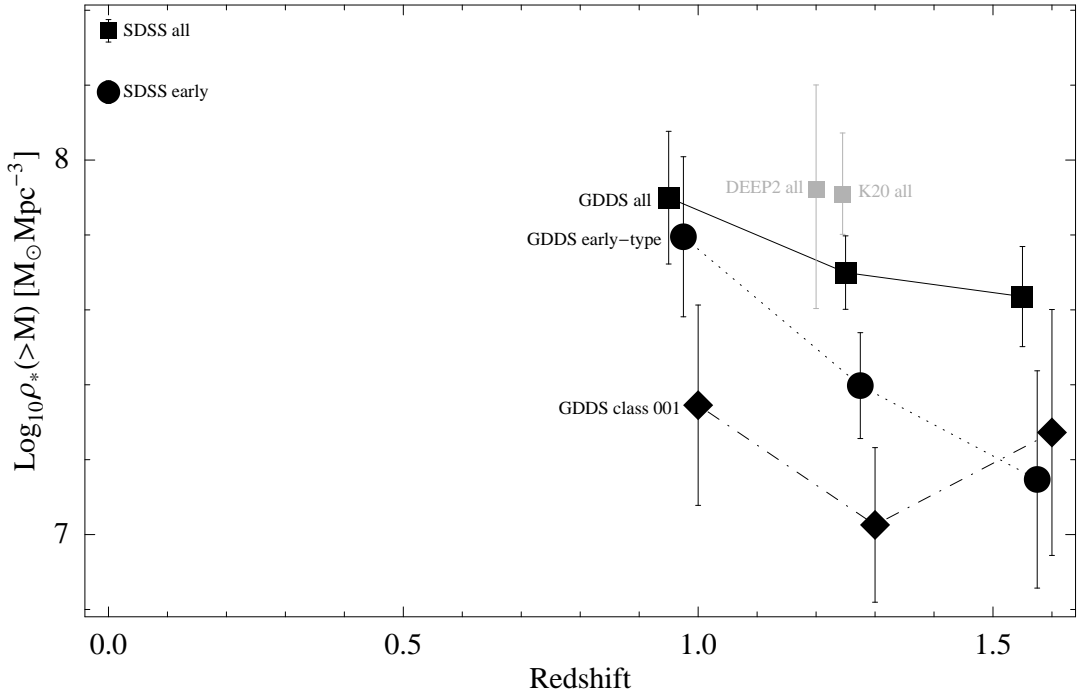} 
\caption{\label{fig:mdf001} 
Mass density functions for subsets of massive galaxies. As described in the text, our stellar mass cut is set at $\log_{10}(M/M_\odot)>10.5$. Plot symbols denote the following subsets:  quiescent (spectral class 001) galaxies [diamonds]; morphologically selected (on the basis of position in the A-G plane) early-type galaxies [circles]; and all galaxies [squares]. The corresponding local stellar mass densities for massive galaxies, taken from the analysis of SDSS observations given by \citet{bel03} and converted to our IMF and mass cut, are also shown. The two points shown in light gray are taken from DEEP2 survey work presented by \citet{bun06}, and from the K20 analysis published in \citet{fon04}}
\end{center}
\end{figure*}

The cumulative mass density function for systems more massive than $\log_{10}(M/M_\odot)=10.5$ is shown in Figure~\ref{fig:mdf001}.   The mass density function shown was computed
in three redshift bins ($0.8<z<1.1$, $1.1<z<1.4$, and $1.4<z<1.7$). The corresponding error
bars  were estimated by a bootstrap analysis.  Plot symbols in Figure~\ref{fig:mdf001} denote the following subsets of galaxies:  (1) Spectroscopically quiescent objects (spectral class 001) shown as diamonds; (2) morphologically selected (on the basis of position in the A-G plane) early-type galaxies shown as circles; and (3) all galaxies shown as squares. At the $z=0$ position we also plot the corresponding local stellar mass densities, based on the analysis of SDSS observations given by \citet{bel03} and converted to our IMF, and incorporating our mass cut.  (Note however that the local early-type point is based on color selection, rather than on morphological classification; this point is discussed further below). The cumulative mass functions for all galaxy types from \citet{fon04} and \citet{bun06} are shown in gray.

The most striking aspect of Figure~\ref{fig:mdf001} is that the cumulative mass locked up in morphologically-selected early-type galaxies appears to be increasing with cosmic time substantially more quickly than the global rate.  The cumulative stellar mass locked up in massive galaxies as a whole is increasing slowly by comparison, changing by only about 0.3 dex over the redshift range probed, in agreement with the trends shown in Paper~III. (This is reassuring, because our ACS imaging observations cover only 55\% of the area in the complete GDDS). The cumulative stellar mass in spectroscopically  quiescent systems is consistent with being flat, although the error bars are large, and we cannot rule out that the spectroscopically quiescent systems are evolving at a similar rate to that of the total galaxy population.

\begin{figure*}[htbp]
\begin{center}
\includegraphics[width=6.5in]{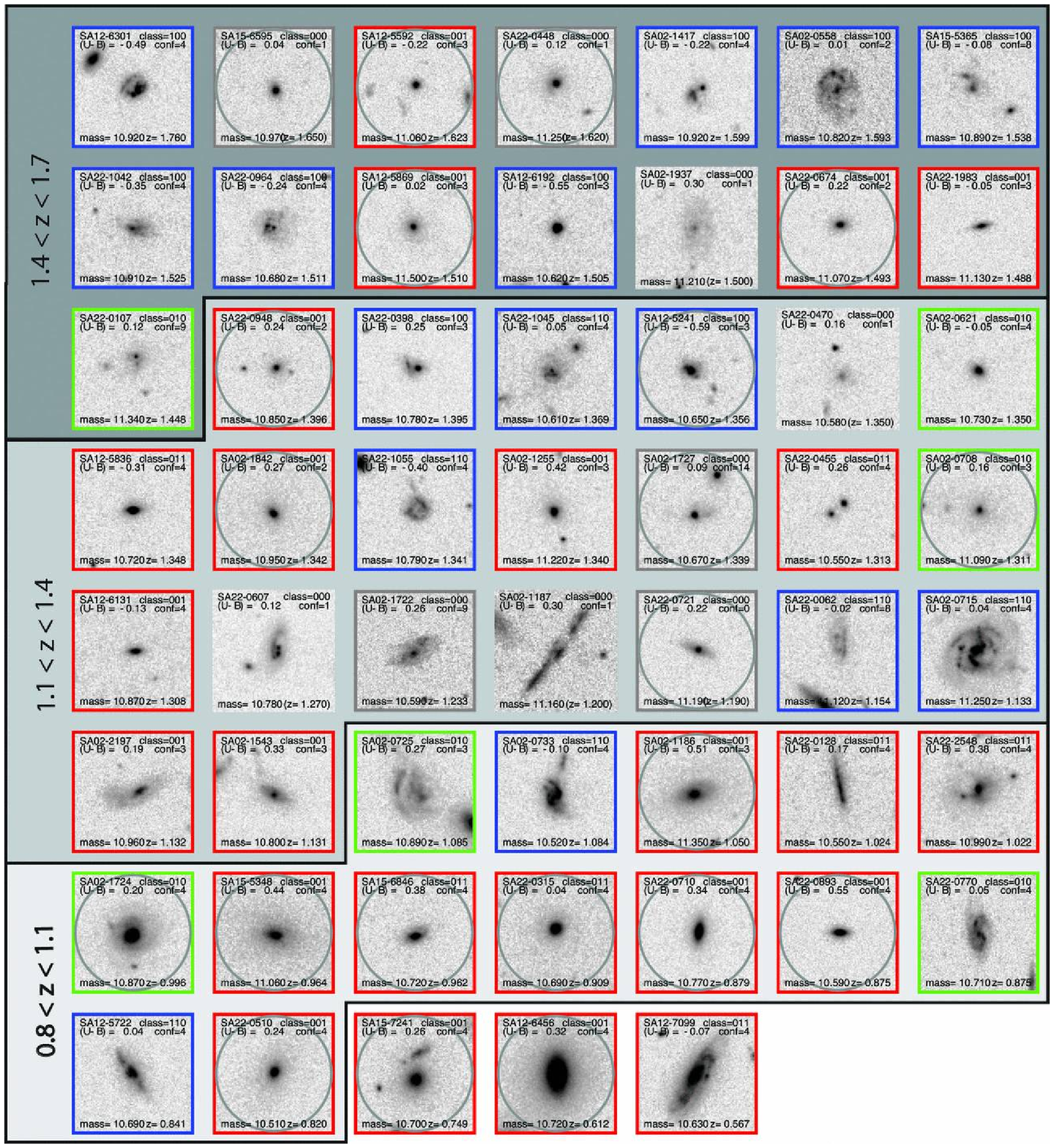} 
\caption{ 
\label{fig:massivePostageStampsSortedByRedshift}
Postage-stamp images showing the morphologies of the 54
galaxies in our sample with $\log_{10}(M/M_\odot)>10.5$, 
sorted in order of decreasing redshift. Early-type galaxies are circled. Each image
is 5 arcsec X 5 arcsec in size, and labeled with the galaxy's ID number, spectroscopic
classification, redshift confidence class, rest-frame (U-B) color, redshift, and stellar
mass inferred from our best-fit model. Objects without high-confidence spectroscopic
redshifts have their redshifts labeled in parentheses. The border of each galaxy image 
is colored according to the galaxy's spectroscopic classification, as described in the caption
to Figure~\ref{fig:ga}. Gray regions surrounding
groups of postage stamps indicate which of three broad
redshift bins 
the objects fall within. These bins are used to calculate the
cumulative stellar mass function described in the text.}
\end{center}
\end{figure*}

In our highest redshift bin the fraction of the total stellar mass contained in massive early-type galaxies is similar to that contained by spectroscopically quiescent systems (around 30\% of the total). In our lowest redshift bin over twice as much mass is contained in early-type galaxies as is contained in quiescent systems --- in fact, early-type systems in our lowest redshift bin contain about 80\% 
of all the stellar mass locked up in galaxies with $\log_{10}(M/M_\odot)>10.5$. This is consistent (within the considerable error bars) to the corresponding value in the local Universe (69\%), which can be obtained from Figure~\ref{fig:mdf001} by simply dividing the values shown for the two points at $z=0$. Figure~\ref{fig:mdf001} therefore
suggests that over the redshift range probed by the GDDS we are witnessing the formation
of early-type galaxies as the dominant members of the high-mass end of the galaxy population,
the same position they maintain up to the present day. The result from the mass function is
in entirely consistent with the visual impression one obtains by simply plotting the morphologies
of massive galaxies in our sample as a function of redshift, as shown in 
Figure~\ref{fig:massivePostageStampsSortedByRedshift}. The postage stamp images
in this figure are shown in order of decreasing redshift, with early-type systems
(based on our quantitative definition) circled. The steady increase with cosmic epoch in the fraction of
early-type systems as a fraction of the massive galaxy population is rather striking. It
seems fairly plausible to conclude that at at lowest redshifts probed by the 
GDDS, early-type galaxies comprise a similar
fraction of the massive galaxy population as is seen locally.

\section{DISCUSSION}

In the local Universe early-type galaxies are nearly always associated with 
quiescent stellar populations, so the steep evolution of early-types in the cumulative mass
function shown in Figure~\ref{fig:mdf001} might be
viewed as a surprise given the relatively shallow evolution in totally quiescent galaxies. A plausible
explanation for this mismatch is offered by the possibility that
morphological transformations and stellar mass assembly operate on different time scales.
For example, it has long been argued \citep{bau96,kau98,del04} that 
the bulk of the stars in quiescent early-types galaxies might have formed before these stars
were organized into 
spheroids, in which case the progenitors of such galaxies might be found in other morphological
types. On the other hand, it might also be argued that
the signatures
of star-formation (e.g. blue colors and emission lines) remain in place for a period of time
subsequent to the formation of an early-type galaxy, in which case the
progenitors of quiescent early-type systems might still resemble early-types, but with 
blue colors and/or emission lines.  We suspect that both scenarios are at play
in driving the steep evolution in the cumulative mass function of early-type
galaxies shown in Figures~\ref{fig:mdf001} and \ref{fig:massivePostageStampsSortedByRedshift}.

\begin{figure*}[htbp]
\begin{center}
\includegraphics[width=6.5in]{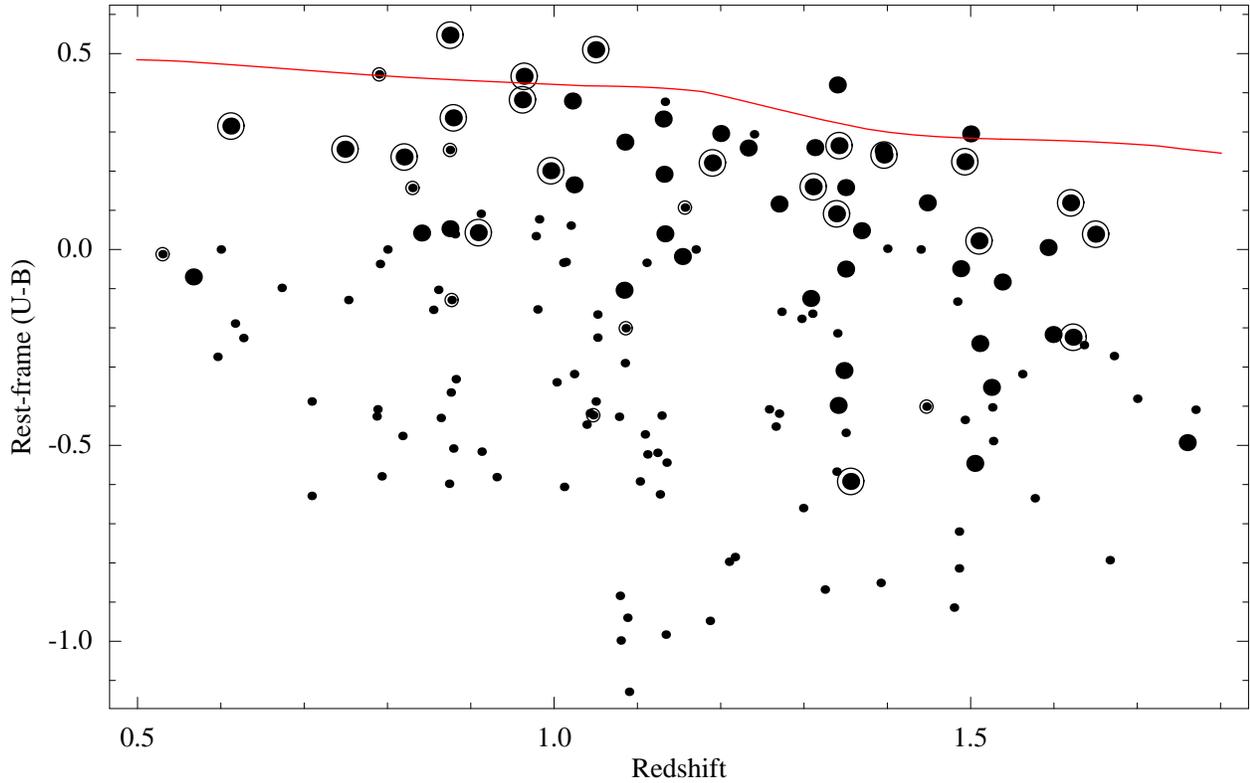} 
\caption{\label{fig:colorz} 
Rest-frame $(U-B)$ color of our sample, plotted as a function
of redshift. Galaxies with $\log_{10}(M/M_\odot)>10.5$ are shown with large symbols.
Morphologically classified early-type galaxies are circled. A fraction of the other red galaxies are massive are highly-reddened star-forming galaxies (Noll et al. 2006). The
line shown near the top of the figure is the evolutionary track of a massive
instantaneous starburst forming all of its stars at $z=3$. See text for details.}
\end{center}
\end{figure*}


To further explore the link between morphology and star-formation history, Figure~\ref{fig:colorz} shows
the rest-frame $(U-B)$ color of our sample, plotted as a function
of redshift. The rest-frame colors of our galaxies were computed using the best-fit spectral energy distribution templates
used to construct our mass functions.
The
line shown near the top of the figure is the evolutionary track of an
instantaneous starburst forming all of its stars at $z=3$. Morphologically-classified early-type galaxies are circled.  A number of interesting points emerge from this figure. As noted earlier,
most morphologically-defined massive early-type galaxies
are very red, but the figure shows that a number of massive early-type galaxies with blue
colors, probably from young stellar populations (as they do not exhibit AGN features in their spectra).
The existence of these objects should come as no surprise, since studies of internal color variations in early type galaxies have shown that a significant fraction of early-type galaxies at $z\sim1$ have blue cores \citep{abr99,men04,pap05}. Furthermore, such systems emerge from any scenario in which early-type galaxies are being built up over the redshift range of our survey with minor bursts of star formation occurring in time-scales longer than the stellar relaxation time. It is interesting to note that at high redshifts the red-end of the galaxy distribution lies consistently to the blue of the passive evolution model, and that at low redshifts most blue early-type systems are not
very massive --- presumably both these effects are manifestations of cosmic downsizing \citep{cow96,kod04,gla04,fon04,bau05,jun05,fab05,bun06}.
We speculate that most of the `blue' early-type galaxies are ultimately destined to turn into the quiescent early-type galaxies of intermediate mass. We also note that Figures~\ref{fig:ga} and 
\ref{fig:massivePostageStampsSortedByRedshift} both show
that the abundance of this class of object
would increase significantly if we were less stringent in requiring that early-type galaxies have low asymmetry, since inspection
of these figures shows that a number of high Gini systems with high asymmetry have not been included in our sample
of early-type galaxies. The asymmetric component of these massive objects is near their nuclei, and we suspect they are late-stage
mergers.
A number of such systems exhibit quiescent spectra, and these would
seem to be 
be good candidates for ``dry merging'' \citep{van05}, though they may
also be systems with
high dust contamination. In any case, these objects are heavily outnumbered
by star-forming systems. 
Figure~\ref{fig:colorz}  also shows that a sizeable fraction of galaxies with very red rest-frame colors are {\em not} early-types. These objects comprise only 15\% of the red galaxies at $0.7<z<1.0$, but 
contamination rises quickly with redshift, and by 
$z\sim 1.5$ about half the objects with red rest-frame colors are not early-types.
While the existence of these objects is not a surprise --- a number of recent papers have
shown that color selection used to define early-type galaxy samples
results in a mixed bag of galaxies \citep{yan03,mou04} --- the
rapid change in the contamination rate as a function redshift is rather striking.

\begin{figure*}[htbp]
\begin{center}
\includegraphics[width=6.5in]{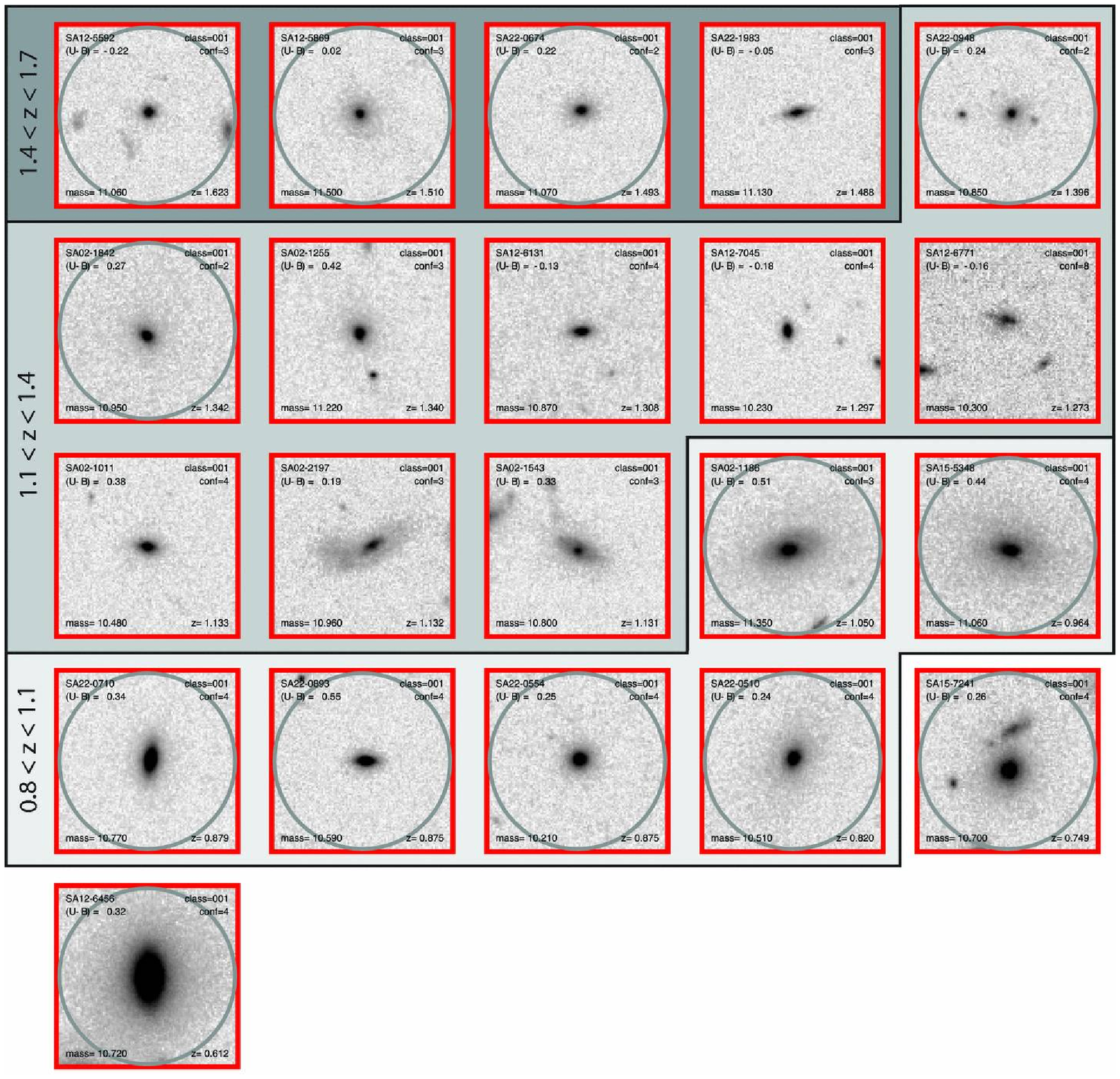} 
\caption{\label{fig:001} 
5 arcsec by 5 arcsec images of the 21 galaxies in our
ACS imaging sample with spectral classifications corresponding to quiescent
stellar populations (class 001). Early-type galaxies are
circled. Gray regions surrounding
groups of postage stamps indicate which of three broad
redshift bins 
the objects fall within. These bins are used to calculate a
cumulative mass function, as described in \S\ref{sec:mfunc}. See text for details.
}
\end{center}
\end{figure*}

\begin{figure*}[htbp]
\begin{center}
\includegraphics[width=6.0in]{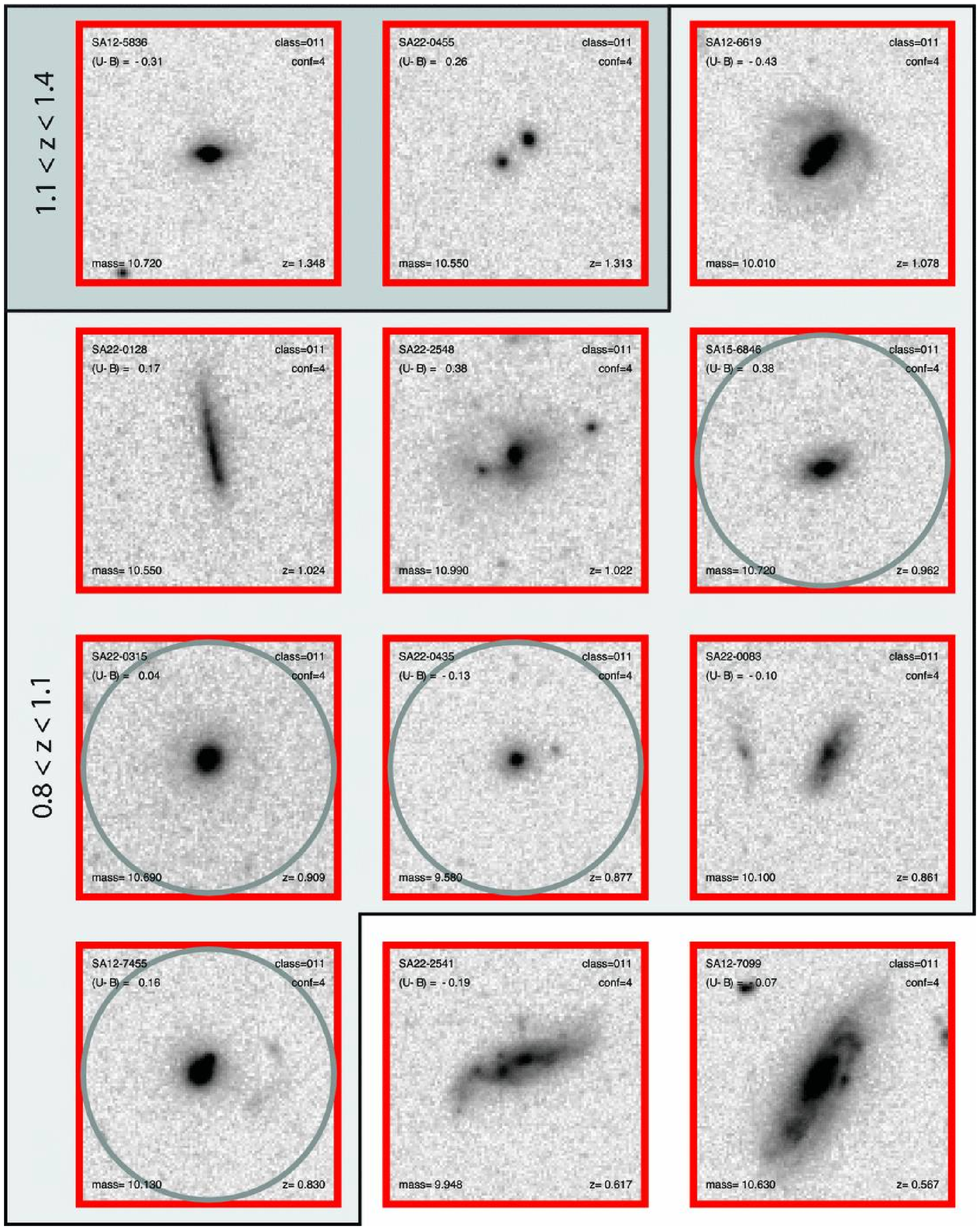} 
\caption{\label{fig:011} 
5 arcsec by 5 arcsec images showing the morphologies of the 12 galaxies in our
imaging sample with spectral classifications corresponding to a dominant old
stellar population augmented by low-level star formation (class 011). Early-type galaxies are
circled. Gray regions surrounding
groups of postage stamps indicate which of three broad
redshift bins 
the objects fall within. These bins are used to calculate a
cumulative mass function, as described in \S\ref{sec:mfunc}.
}
\end{center}
\end{figure*}

Since color-selected samples cannot be used to define samples of early-type galaxies
without significant contamination, it is interesting to consider whether better samples
of early-types might be constructed using spectroscopy.  Figure~\ref{fig:001} shows the F814W images
of the 21 galaxies classified as purely quiescent (class `001' in the taxonomy of Paper~I) in our 
ACS sample. About 2/3 of the sample (13/21) are early-types using our automated classifier. About half the remainder are objects that would be classified as early-types if our asymmetry cut were slightly less rigorous, and on the whole the sample appears fairly homogeneous. We conclude that a fairly good sample of early-type galaxies could be defined simply by selecting purely quiescent systems. However, Figure~\ref{fig:011} makes it clear that care would have to be taken to exclude objects showing even small amounts of star-formation. This figure shows a montage of the 12 galaxies in our
imaging sample with spectral classifications corresponding to a dominant old
stellar population contaminated by low-level star formation (class `011' in the taxonomy of Paper~I).
It is obvious that these objects have a wider range of morphologies than those with the `001' (quiescent) spectral class and a higher fraction of disks, and 2/3 of these objects lie outside the region in
the A-G plane used to isolate early-type
systems. Approximately 50\% of the old + weak/truncated star formation objects are disks and many show prominent HII regions.  The remaining objects are compact and have early-type morphologies. We can only speculate on the ultimate destiny of the large disk galaxies
in this sample. Several of these objects have prominent bulges, and fading of their disks might bring them rather close to meeting our criterial for early-type galaxies. It is tempting to identify these with fairly recent merger events that will relax to the passive and early-type systems at $z < 1$. Indeed these objects are closely related to the massive post-starburst systems described in \citet{leb06} (Paper VI). They appear to have undergone significant star formation episodes at $z \ge 1.5$ and, barring further star burst episodes, will appear as passive systems on the red sequence by $z \sim 0.8-1$.  (We note that spectroscopy of color-selected samples -- \citet{yan03} --  have also revealed luminous disk galaxies with red colors and spectra dominated by old stars.) 

Our results paint a picture of the {\em morphological assembly} of many massive early-type galaxies over the redshift range $1 < z < 2$, but it is important not to take this conclusion too far. Large area surveys (Brown et al. 2006) clearly indicate that the most massive systems have formation redshifts, both stellar and dynamical, well above the redshift limit of our present sample. The age analysis carried out in Paper IV and similar analyses of other samples \citep[e.g.][]{cim04,hea04,jim06} suggest that these objects formed at $z > 3$ on average and in some cases at considerably higher redshifts. As we will show in a subsequent paper, the rest-frame R-band morphologies of the passive galaxies with $z > 1.3$ are consistent with early dynamical formation as well. These results are another manifestation of the popular down-sizing paradigm in which massive galaxies form early and less massive systems are assembled at later times \citep{cow96,jun05}. 
At the low-redshift end of our sample, the convergence in the fractional mass density in massive early-type galaxies (relative to the total mass density at the same redshift) to something similar to the value in the SDSS indicates that the epoch of massive early-type formation may be drawing to a close by $z\sim 1$ (a conclusion consistent with the high post-starburst fraction described in Paper VI).

The offset in the mass density contained in the lowest redshift GDDS point in Figure~\ref{fig:mdf001}
and the local data point is
rather striking, and this {\em may} indicate that whatever process is driving the process of massive galaxy assembly at $z<1$ is operating in a way which conserves the overall fraction of early-types even as it builds the total stellar mass in the galaxy population from $z=0.7$ to $z=0$.  However, we remain a little skeptical about this conclusion. The most robust statements that can be made from  Figure~\ref{fig:mdf001} are based on {\em relative measurements internal to each dataset}.  Comparing the high-redshift GDDS points to their low-redshift counterparts in Figure~\ref{fig:mdf001} is dangerous, even though we chosen to plot them together in the same diagram. \citet{bel03} note that 
there is an at least $\sim30\%$ systematic uncertainty in the local estimate, and the uncertainty in the high-redshift points is likely to be at least this big. Another source of potential bias is that
\citet{bel03} define their early-type sample on the basis of color, rather than on the basis of
morphology. Much of the offset between the cumulative stellar mass locked in
early type at $z=0$ and $z>0.7$ may be due to systematic differences in the way 
samples of high-redshift and low-redshift early-type galaxies are constructed --- once again, we
emphasize that the
trends seen internally to either data set are more reliable.

\section{CONCLUSION}

We have described an improved methodology for morphologically
classifying galaxies, based on `quasi-Petrosian' image segmentation. This
results in more robust measurements of galaxy properties at high redshifts.
This methodology has been applied to define a sample of early-type
galaxies in HST ACS images of the Gemini Deep Deep Survey fields. 
Using this sample, we computed the cumulative stellar mass functions 
of morphologically-segregated subsets of galaxies with 
$\log_{10}(M/M_\odot)>10.5$. Although our error
bars are large, we find striking evidence for 
evolution in the fraction of stellar mass locked up in massive
early-type galaxies over the redshift range $0.7<z<1.7$. 
This redshift range corresponds to that over which massive
early-type galaxies morphologically assemble,  and 
over which the strong color-morphology correlations
seen in the local Universe begin to fall into place.

\acknowledgments
\noindent{\em Acknowledgments}

\noindent 
This paper is based on observations obtained at the Gemini
Observatory, which is operated by the Association of Universities for
Research in Astronomy, Inc., under a cooperative agreement with the
NSF on behalf of the Gemini partnership: the National Science
Foundation (United States), the Particle Physics and Astronomy
Research Council (United Kingdom), the National Research Council
(Canada), CONICYT (Chile), the Australian Research Council
(Australia), CNPq (Brazil) and CONICET (Argentina).

Based on observations made with the NASA/ESA Hubble Space Telescope, obtained at the Space Telescope Science Institute, which is operated by the Association of Universities for Research in Astronomy, Inc., under NASA contract NAS 5-26555. These observations are associated with program \#9760. Support for program \#9760 was provided by NASA through a grant from the Space Telescope Science Institute, which is operated by the Association of Universities for Research in Astronomy, Inc., under NASA contract NAS 5-26555.

RGA thanks 
NSERC, the Government of Ontario, and the Canada Foundation for
Innovation
for funding provided by an E. W. R. Steacie Memorial Fellowship.  
KG \& SS acknowledge generous funding from the David and
Lucille Packard Foundation. H.-W.C. acknowledges support by NASA
through a Hubble Fellowship grant HF-01147.01A from the Space
Telescope Science Institute, which is operated by the Association of
Universities for Research in Astronomy, Incorporated, under NASA
contract NAS5-26555. DLB wishes to thank the Centre National d'Etudes Spatiales for its support.

\bigskip
\appendix
\bigskip
\centerline{\Large\bf Appendices}

\section{ROBUSTNESS OF THE QUASI-PETROSIAN GINI COEFFICIENT AS A GALAXY CLASSIFIER}
\label{sec:app}

In this Appendix we will investigate the robustness of our galaxy classification
methodology. As described in Section~4, our classifications are based on
a galaxy's position in the Asymmetry vs. Gini diagram. For purposes of
computing the mass function of early-type galaxies, we have shown in the text that the 
Gini coefficient
plays a much more significant role than does asymmetry. In fact, the main
conclusions of this paper would remain unchanged if asymmetry were neglected completely. 
Therefore our
aim in this appendix is to demonstrate that the Gini coefficient measured
in a single red band (F814W) can be used as a
simple and very robust classifier of early-type galaxy morphology out to $z\sim2$. 
We emphasize at the outset that this claim is true {\em only} when data is deep enough to 
allow
quasi-Petrosian isophotes to be used to calculate the Gini coefficient,
as described in the text.

\begin{figure*}[htbp]
\begin{center}
\includegraphics[width=4.0in]{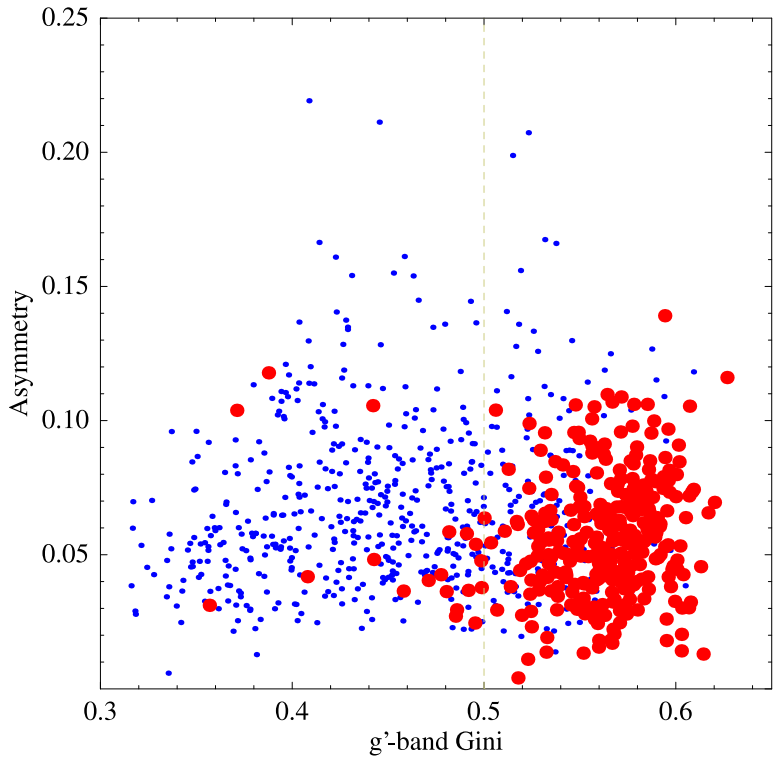} 
\caption{\label{fig:sdssag} 
Asymmetry vs. Gini coefficient for 800 nearby galaxies
observed in $g$'-band as part of the Sloan Digital Sky Survey. Quasi-Petrosian
thresholds were used when measuring both quantities, as described in the
text.
Galaxies classified visually as being early-type
systems are shown in red. The vertical line denotes the $G=0.5$
cut used to distinguish early-type galaxies from all other
systems.
}
\end{center}
\end{figure*}

Our analysis of the robustness of the quasi-Petrosian Gini coefficient
is based on measurements of this quantity obtained through
$u$', $g$', $r$' and  $i$' filters for 800 nearby galaxies in
the
Sloan Digital Sky Survey (SDSS).  
This is a subset of the morphological sample analyzed by Nair et al. (2007). 
We refer readers to that paper for additional details. Nair
et al. (2007) will also present detailed 
comparisons of classifications made using quasi-Petrosian
Gini coefficients to those made using other methods. 
Our ACS imaging sample has a signal-to-noise ratio floor of 100 (only
a single galaxy in our ACS imaging sample has has a signal-to-noise ratio below this) so
our analysis will be restricted to galaxies with a $u$'-band signal-to-noise ratio $>100$.

Figure~\ref{fig:sdssag} shows the Asymmetry vs. Gini coefficient diagram for $g$'-band
imaging of our SDSS sample. Galaxies classified visually by one of us (Preethi Nair) 
as being early-type
systems are shown in red. The vertical line denotes the $G=0.5$
cut used to distinguish early-type galaxies from all other
systems. It is seen that this simple bifurcation of the A-G plane into
two parts does a rather good job of distinguishing early-type galaxies
from all other systems. As expected, there is some leakage of visually-classified 
early-type galaxies to regions
of the diagram with $G<0.5$, and some leakage in the other direction 
(non-early-types into the $G>0.5$). These systems are almost all
early-type spirals whose visual classifications are known (based on 
comparisons of visual classifications between observers, e.g. Naim et al. 1995) to 
be ambiguous
(e.g. Sa vs. S0/a vs. S0 systems). 

\begin{figure*}[htbp]
\begin{center}
\includegraphics[width=6.0in]{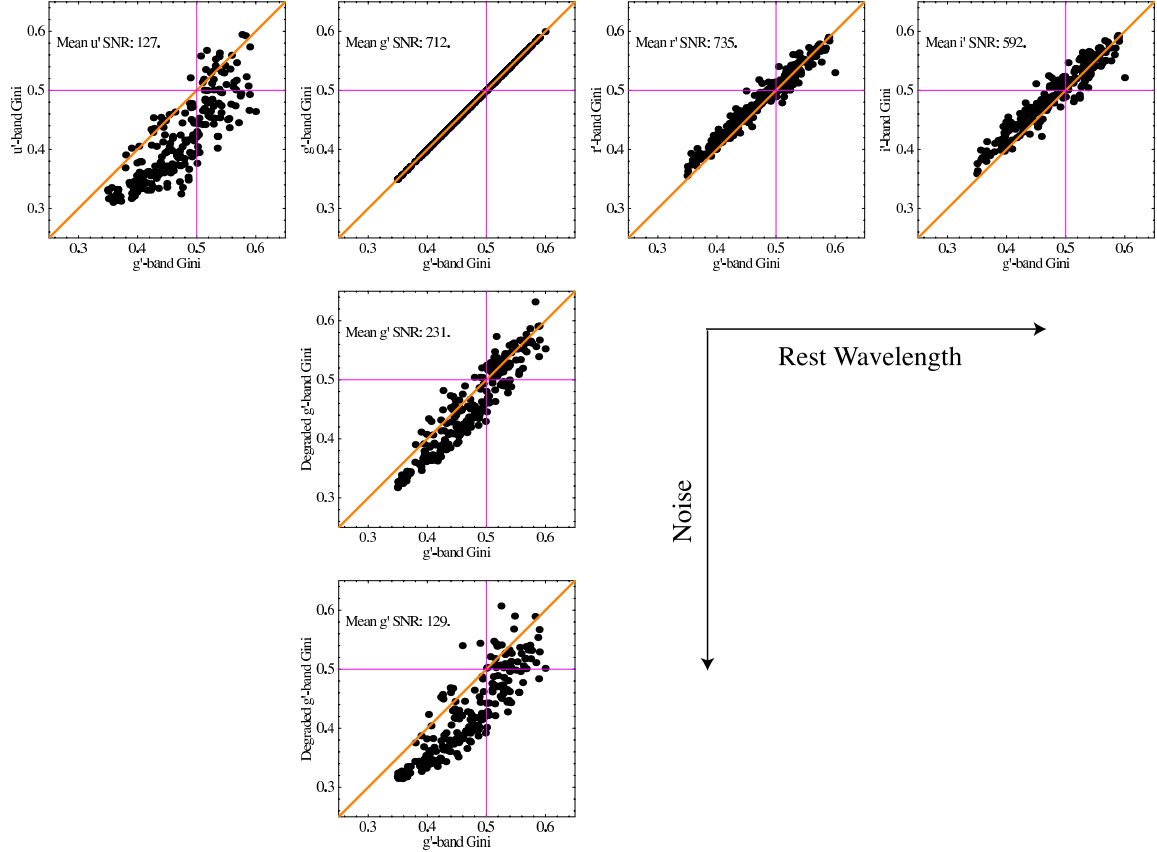} 
\caption{\label{fig:sdssall}
Analysis of the robustness of the quasi-Petrosian Gini coefficient,
based on measurements of this quantity obtained through
$u$', $g$', $r$' and  $i$' filters for 800 nearby galaxies in
the
Sloan Digital Sky Survey (SDSS). The x-axis of each
panel in this figure is keyed to the $g$'-band Gini coefficient measured for each galaxy
in this sample.  
The top row shows the Gini coefficients obtained at different wavelengths (through $u$', $g$', 
and $i$' filters) plotted against the Gini 
coefficients in $g$'-band. The diagonal line shown in each panel has a slope of unity and intersects the origin, and therefore corresponds to no change in Gini as a function of wavelength. Vertical and horizontal lines shown in each panel at Gini coefficients of 0.5 correspond to our proposed cutoff for early and late-type galaxies. Changes in the Gini coefficient are remarkably small over the entire wavelength range spanning the $g$' through $i$' filter set. While Gini seems to change significantly when comparing $u$'-band to $i$'-band in the SDSS, this is easily shown to an artifact introduced by the low signal-to-noise in the SDSS $u$'-band data (which is typically much shallower than that obtained in other bands). To explore the sensitivity of the Gini coefficient to noise, the base of the `T' in the figure shows how the Gini coefficient of  $g$'-band sample changes when noise is added to the images. The mean signal-to-noise ratio of the noise-degraded $g$'-band data in middle panel is similar to the mean signal-to-noise ratio for galaxies at $z>1.2$ in our sample. The mean signal-to-noise of the bottom panel is similar to that of galaxies with the poorest data in our high-redshift sample, and also to the mean signal-to-noise of the data in the SDSS $u$'-band sample. The Gini coefficients of noise-degraded $g$'-images and $u$'-band images are compared explicitly in the next figure.
}
\end{center}
\end{figure*}

The $G=0.5$ cut evidently 
does a good job of distinguishing high signal-to-noise $g$'-band images of
SDSS early-type galaxies from the rest of the
galaxy population, but how robust is this cut to changes in rest wavelength of
observation, and to decreasing signal-to-noise? Figure~\ref{fig:sdssall} is
an attempt to address this question.
The top row of Figure~\ref{fig:sdssall} shows the Gini coefficients
of our SDSS reference sample 
measured at different wavelengths (using $u$', $g$', 
and $i$' filters) plotted against the Gini 
coefficients in $g$'-band. The diagonal line shown in each panel delineates 
a perfect mapping between the two parameters, {\em i.e.}
no change in Gini as a function of wavelength. The quasi-Petrosian Gini
coefficient is seen to be remarkably robust to changes in rest wavelength 
when measured from data with reasonably high signal-to-noise.
The Gini coefficient measured in $g$'-band barely changes when measured
in $r$'-band and $i$'-band, and for early-type
galaxies in particular the changes are hardly bigger than the random 
measurement errors (Nair et al. 2007). 
Measurements of the $u$'-band Gini coefficient probe
wavelengths blueward of the 4000\AA~break, and our
na\"ive expectation was that Gini coefficients {\em should} be significantly different
in $u$'-band when compared with measurements made at wavelengths 
redward of the break.  At first inspection Figure~\ref{fig:sdssall} does seem to 
show this, in the form of systematic offsets between the $g$'-band and $u$'-band
images. However, a more careful inspection shows that the trends seen
are almost certainly due to the the low signal-to-noise ratio of the $u$'-band
data, instead of being due to systematic differences in the intrinsic $g$'-band
and $u$'-band images of galaxies. 

Each panel in Figure~\ref{fig:sdssall} records the mean signal-to-noise level
of the sample being plotted. It is seen that
the SDSS $u$'-band data
has a mean signal-to-noise ratio of 
$\sim 130$, which is a factor two to three lower than the mean signal-to-noise ratios
of our ACS images, and about a factor of five lower than the signal-to-noise ratios
of the corresponding 
SDSS $g$', $r$', and $i$' images. 
It is clear that these differing signal-to-noise ratios need to
be accounted
for before comparing the $u$'-band Gini coefficients of the data obtained at different wavelengths. 
This is a crucial point, because the ACS F814W filter
begins to probe blueward of the 4000\AA~ break at $z>1$, where the
bulk of our data lies. Since
the SDSS $u$'-band filter data would
seem to provide the best match to the 
rest-frame wavelengths being probed by much of our ACS data, understanding the
reason why the
Gini coefficients of $u$'-band data shown in
Figure~\ref{fig:sdssall} appear be systematically lower than those 
measured at longer wavelengths is of special significance.
The columnar
base of the `T' in Figure~\ref{fig:sdssall} is an attempt to understand the $u$'-band data
by showing 
how the Gini coefficient of  $g$'-band sample changes when noise is added to the images to
lower the mean signal-to-noise ratios of the samples plotted. The mean signal-to-noise ratio of the 
noise-degraded $g$'-band data in middle panel is similar to the mean signal-to-noise ratio for 
galaxies at $z>1.2$ in our sample. The mean signal-to-noise ratio of the 
lowest panel is similar to that of the data in the SDSS $u$'-band sample. 
The strong resemblance between this panel and the $u$'-band panel at the top
left of the Figure suggests that the apparent offset in the Gini coefficients of the high signal-to-noise
$g$'-band data and the low signal-to-noise $u$'-band data is mostly due to poor signal in the latter,
and not due to morphological K-corrections.

\begin{figure*}[htbp]
\begin{center}
\includegraphics[width=6.0in]{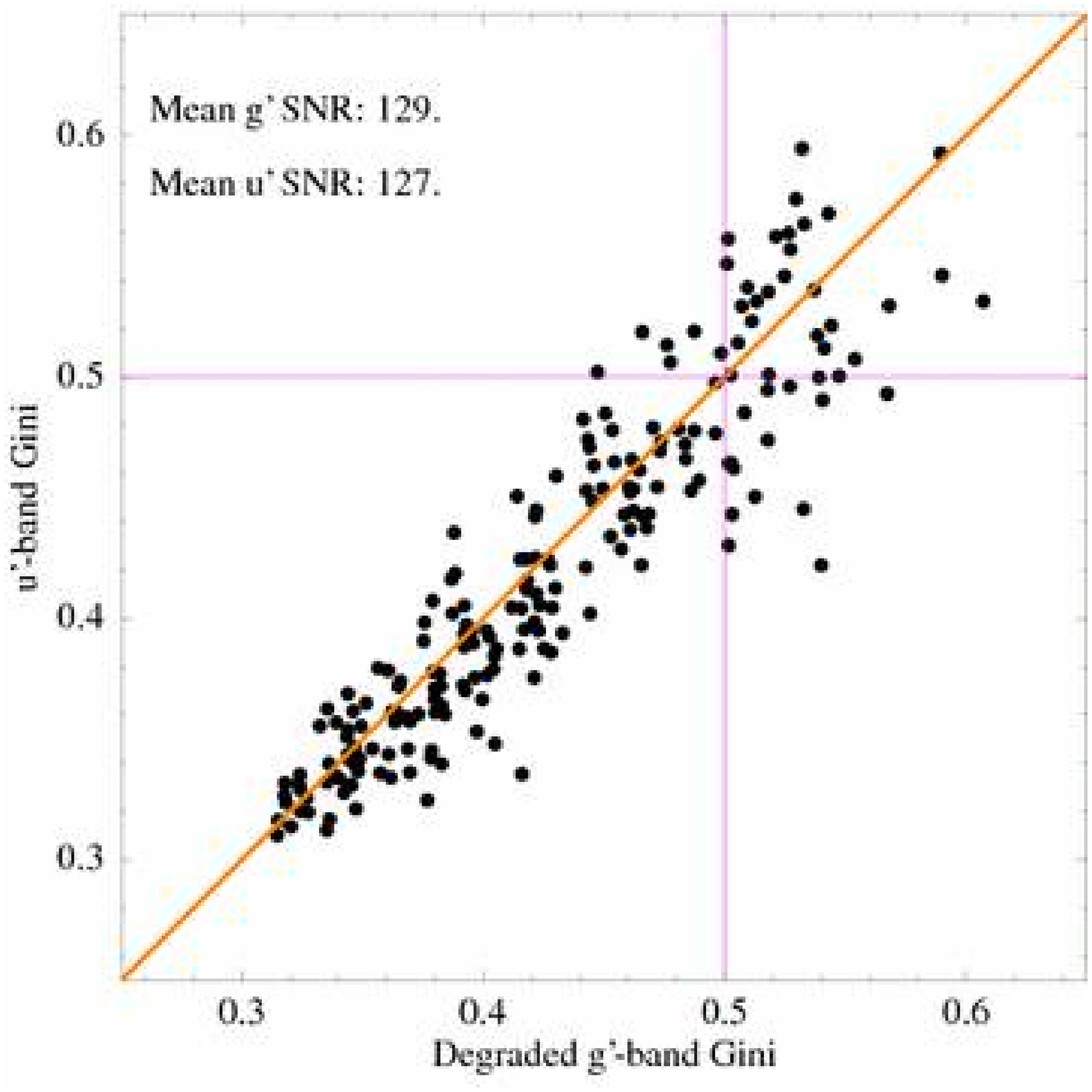} 
\caption{\label{fig:sdssug} 
Gini coefficients of $u$'-band galaxies in the SDSS compared
to Gini coefficients of corresponding 
noise-degraded $g$'-band images. The mean signal-to-noise
ratios of the samples have been harmonized to be nearly
identical ($\sim130$). The scatter is consistent with the
random measurement
error at these low signal-to-noise levels (see Nair et al. 2007 for details).
As described in the text, when a 
quasi-Petrosian formalism is adopted for making the
measurements, an individual
galaxy's Gini coefficient is nearly independent of its
wavelength of observation.
}
\end{center}
\end{figure*}

A more explicit comparison of $u$'-band and noise-degraded
$g$'-band Gini coefficients of our local calibration sample
is shown in Figure~\ref{fig:sdssug}. The mean signal-to-noise
ratio of the $g$'-band sample has been harmonized to be nearly
identical that that of the $u$'-band sample. Within
the scatter introduced by random measurement
error, Gini
coefficient measurements at $g$'-band and $u$'-band
are identical,
even
though these bands straddle the 4000\AA~break (at which point visual
morphologies of galaxies can appear to make large changes). This
highlights the strongest benefit of using an
abstract quantity like the
quasi-Petrosian Gini coefficient as measure of morphology. This is also
the essence of our contention, made repeatedly in the main text of this paper, that
if quasi-Petrosian Gini coefficients are used to quantify morphology, then a basic
`cure' for the
deleterious effects of morphological $K$-corrections is simply to obtain deeper data.

\begin{figure*}[htbp]
\begin{center}
\includegraphics[width=6.0in]{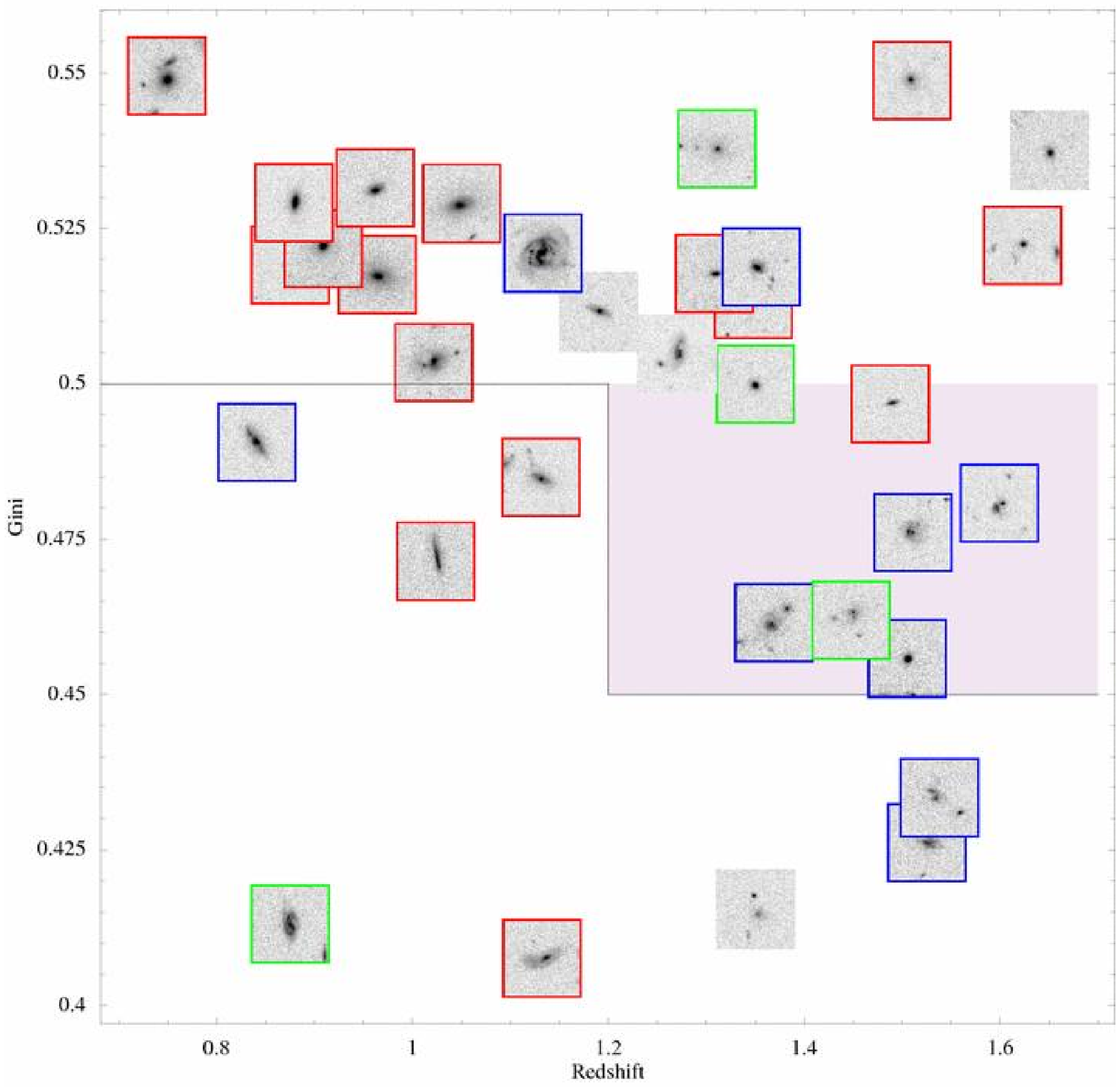} 
\caption{\label{fig:gzzoom} 
An object-by-object comparison of Gini coefficient
vs. redshift for galaxies near the border of the $G=0.5$
cutoff threshold used to discriminate between early-type
galaxies and all other systems. Objects in the 
gray region shown  are worth looking at individually
because they are at redshifts where F814W imaging
is probing blueward of the 4000\AA~break and also
because
they are
near enough to the $G=0.5$ cutoff that they could
be brought below it at very low signal-to-noise. 
However, all these objects are seen to be high signal-to-noise
systems, and only one object has a visual
morphology even remotely consistent with that of 
an early-type galaxy. As in Fig. 5, the border of each postage stamp is
is colored according to its spectral classification based on the system described
in Paper I. See caption to Fig.5 for details.}
\end{center}
\end{figure*}

We conclude this appendix with Figure~\ref{fig:gzzoom}, which
shows
an object-by-object postage-stamp image montage
of galaxies near the $G=0.5$
border of the Gini coefficient
vs. redshift diagram. This diagram can be used
to inspect the morphologies of objects near the
cutoff threshold used to discriminate between early-type
galaxies and all other systems in the present
paper. Objects in the 
gray region  shown in Figure~\ref{fig:gzzoom} shown are of
particular interest, because these systems are 
at redshifts where F814W imaging
is probing blueward of the 4000\AA~break, and also
near enough to the $G=0.5$ cutoff that they could
conceivably be brought below it at very low signal-to-noise. 
However, all these objects are seen to be high signal-to-noise
systems, and only one object (with a high star-formation
rate spectrum) has a visual
morphology even remotely consistent with that of 
an early-type galaxy.

\end{document}